\documentclass[11 pt]{article}
\usepackage[utf8]{inputenc}
\usepackage{amssymb,amsmath,amsfonts}
\usepackage{bbm}
\usepackage{times}
\usepackage{graphicx}
\usepackage{subfigure}
\usepackage{color}
\usepackage{multirow}
\usepackage{float}
\usepackage{rotating}
\usepackage{bbm}
\usepackage{enumitem}
\usepackage{caption}
\usepackage{natbib}
\usepackage{hyperref}

\newcommand{\mba}{\mathbf{a}}
\newcommand{\mbf}{\mathbf{f}}
\newcommand{\mbF}{\mathbf{F}}
\newcommand{\mbg}{\mathbf{g}}
\newcommand{\mbh}{\mathbf{h}}
\newcommand{\mbj}{\mathbf{j}}
\newcommand{\mbx}{\mathbf{x}}
\newcommand{\mby}{\mathbf{y}}
\newcommand{\mbz}{\mathbf{z}}
\newcommand{\mbp}{\mathbf{p}}
\newcommand{\mbu}{\mathbf{u}}

\usepackage{xcolor}

\begin{document}


{\LARGE Variational design of sensory feedback for powerstroke-recovery systems}

\ \\
{\bf \large Zhuojun Yu}\\
zhuojun.yu@case.edu\\
Department of Mathematics, Applied Mathematics, and Statistics, Case Western Reserve University, Cleveland, OH 44106, USA\\
{\bf \large Peter J.~Thomas}\\
pjthomas@case.edu\\
Department of Mathematics, Applied Mathematics, and Statistics, Department of Biology, Department of Electrical, Control and Systems Engineering, Case Western Reserve University, Cleveland, OH 44106, USA

Although the raison d'etre of the brain is the survival of the body, there are relatively few theoretical studies of closed-loop rhythmic motor control systems.  
In this paper we provide a unified framework, based on variational analysis, for investigating the dual goals of performance and robustness in powerstroke-recovery systems.
To demonstrate our variational method, we augment two previously published closed-loop motor control models by equipping each model with a performance measure based on the rate of progress of the system relative to a spatially extended external substrate -- such as a long strip of seaweed for a feeding task, or progress relative to the ground for a locomotor task.  
The sensitivity measure quantifies the ability of the system to maintain performance in response to external perturbations, such as an applied load.
Motivated by a search for optimal design principles for feedback control achieving the complementary requirements of efficiency and robustness, we discuss the performance-sensitivity patterns of the systems featuring different sensory feedback architectures.
In a paradigmatic half-center oscillator (HCO)-motor system, we observe that the excitation-inhibition property of feedback mechanisms determines the sensitivity pattern while the activation-inactivation property determines the performance pattern.
Moreover, we show that the nonlinearity of the sigmoid activation of feedback signals allows the existence of optimal combinations of performance and sensitivity.
In a detailed hindlimb locomotor system, we find that a force-dependent feedback can simultaneously optimize both performance and robustness, while length-dependent feedback variations result in significant performance-versus-sensitivity tradeoffs.
Thus, this work provides an analytical framework for studying feedback control of oscillations in nonlinear dynamical systems, leading to several insights that have the potential to inform the design of control or rehabilitation systems.  


\section{Introduction}

Physiological systems underlying vital behaviors such as breathing, walking, crawling, and feeding, 
must generate motor rhythms that are not only \emph{efficient}, but also \emph{robust} against changes in operating conditions.
Although central neural circuits have been shown to be capable of producing rhythmic motor outputs in isolation from the periphery \citep{brown1911intrinsic,brown1914nature,harris1985serotonin,pearson1985there,smith1991pre}, the role of sensory feedback should not be underestimated.
Sensory feedback can play a crucial role in stabilizing motor activity in response to unexpected conditions.
For example, modeling work suggests that walking movements can be stably restored after spinal cord injury by enhancing the strengths of the afferent feedback pathways to the spinal central pattern generator (CPG) \citep{markin2010afferent,spardy2011dynamical}.
Feedback control can also improve the performance and efficiency of movements.
For instance, in a model of feeding motor patterns in the marine mollusk \textit{Aplysia californica}, seaweed intake can be increased by strengthening the gain of sensory feedback to a specific motor neural pool \citep{wang2022variational}.

We are interested in understanding how sensory feedback contributes to control and stabilization within a specific class of rhythmic motor behaviors, namely, behaviors in which an animal (or robot) repeatedly engages and disengages with the outside world.
We refer to the phase of the motion during which the animal is in contact with an external substrate as the \emph{power stroke}, and the component during which the animal is disengaged as the \emph{recovery} phase.  
The decomposition of a repetitive movement into powerstroke and recovery applies naturally to many motor control systems, including locomotion \citep{jahn1972locomotion} and swallowing \citep{shaw2015significance}; a similar dynamical structure also appears in mechanical stick-slip systems \citep{galvanetto1999dynamics} as well as abstract two-stroke relaxation oscillators \citep{JelbartWechselberger2020Nonlin}.
In the motor control context, when the animal is in contact with an external substrate or load opposing the motion, we say the animal makes ``progress" (food is consumed, distance is traveled, oxygen is absorbed) relative to the outside world. 
During the recovery phase, the animal disconnects from the external component, and repositions relative to the substrate in order to prepare for the next power stroke.
Consider, for example, the ingestive behavior of \textit{Aplysia} \citep{shaw2015significance,lyttle2017robustness,wang2022variational}. 
When the animal's grasper is closed on a stipe of seaweed, it drags the food into the buccal cavity; meanwhile, the food applies a mechanical load on the grasper.
Then the grasper opens, releasing its grip on the food.
The grasper moves in the absence of the force exerted by the seaweed and returns to the original position to begin the next swallowing cycle.

In this paper, we present a novel analysis of feedback control for powerstroke-recovery systems.
To quantitatively evaluate the behavior of a system controlled by different feedback mechanisms, we measure the sensitivity (or robustness) and performance (or efficiency). 
The complementary objectives of sensitivity and performance have been studied in a variety of motor control systems, from both empirical and theoretical perspectives \citep{lee1996robust,yao1997high,ronsse2008control,hutter2014toward,lyttle2017robustness,sharbafi2020parallel,mo2023slack}.
There are a myriad of ways to interpret performance and robustness used by engineers, biologists, neuroscientists, and applied mathematicians. 
Here we define the \emph{performance} of a powerstroke-recovery system to be the total progress divided by the period of the rhythm (i.e.~the average rate of progress), and the \emph{sensitivity} to be the ability of the system to maintain performance in response to some perturbations on the external condition (i.e.~the derivative of the performance with respect to the perturbation parameter).
As a step towards first-principles--based design of sensory feedback mechanisms, we aim to understand what aspects of sensory feedback contribute to the coexistence of high performance and low sensitivity.  

The ubiquity of powerstroke-recovery systems, and the importance of the dual goals of robustness and efficiency, motivate us to develop analytical tools for systematically studying both quantities simultaneously.
In this work we apply mathematical tools based on \emph{variational analysis} to evaluate the two objectives applicable for any powerstroke-recovery system.
The key quantities in our analysis are the \emph{infinitesimal shape response curve} (iSRC) and \emph{local timing response curve} (lTRC) recently established by \cite{wang2021shape,wang2022variational} and generalized in \cite{yu2022homeostasis,yu2023sensitivity}.
The iSRC describes, to first order, the distortion of an oscillator trajectory -- the \emph{shape response} -- under a sustained perturbation.
In contrast, the lTRC captures the effect of the perturbation on the \emph{timing} of the oscillator trajectory within any defined segments of the trajectory (such as the powerstroke and recovery phases).
Both the iSRC and the lTRC complement the more widely known infinitesimal phase response curve (iPRC), which quantifies the effect of a transient perturbation on the global limit cycle timing \citep{brown2004phase,izhikevich2008phase,ermentrout2010mathematical,schultheiss2012phase,zhang2013phase}.

Based on the iSRC and lTRC approach, we propose a general framework to investigate   robust and efficient control through diverse feedback architectures.
We apply our method to two neuro-mechanical models, each possessing a natural powerstroke-recovery structure but different in their levels of details and perturbations.
The first model is based on an abstract CPG-feedback-biomechanics system introduced in \cite{yu2021dynamical} which studied the relative contributions of feedforward and feedback control (an idea going back to \cite{kuo2002relative}). 
We extend this model to incorporate an externally applied load, enabling us to define quantitative measures of both performance and robustness, as the system alternately grasps and releases the external substrate.
The second model, due to Markin et al.~\citep{markin2010afferent,spardy2011dynamical}, represents a locomotor system with a single-joint limb, and features more detailed CPG circuitry as well as more realistic afferent feedback pathways.
We modify the Markin model so that the limb ``walks" up an ``incline"; modifying the angle of the incline introduces a parametric perturbation that allows us to define performance and robustness.

The activity of sensory feedback pathways is difficult to measure in many physiological systems; for this reason sensory feedback is the ``missing link" for understanding the design and function of many biological motor systems.  
Specifically, in many experimental biological systems, the dynamics of the isolated CPG and the form of the muscle activation in response to descending motor signals is well characterized, while the precise form of the sensory feedback remains unkown.
From a practical perspective, descending motor signals are generally carried by large-diameter axons from which it is easier to record high-quality (high signal-to-noise) traces, relative to ascending sensory signals, which are generally carried by much smaller diameter axons with poor signal-to-noise properties. 

Given a particular specification of central neural circuit, descending output, and biomechanical response elements, but with the precise form of sensory feedback unknown, we can think of the pursuit of performance and/or robustness as an optimization (or dual optimization) problem on the space of sensory feedback functions. 
However, this is an infinite dimensional space of inputs to a highly nonlinear system. 
That is, the mapping from sensory feedback function to system trajectories to system performance and/or robustness is highly nonlinear and possibly non-convex.  
It may have multiple nonequivalent optima.
Therefore, to restrict the problem to a manageable scope, in the specific cases studied here we restrict attention to sensory feedback functions with a prescribed, but plausible form, such as a sigmoid specified by a threshold and slope parameters, or multiple channels with different relative gain parameters, and study the restricted optimization problem there. 
Moreover, for simplicity, in the models we consider here, we restrict attention to sensory feedback that either monotonically increases or monotonically decreases with variables such as muscle length, tension, and/or velocity.

Our analysis of these two systems -- one more abstract and the other more realistic -- illustrates a technical framework for studying performance and sensitivity of powerstroke-recovery motor systems, leading to several insights that have the potential to inform the design of control or rehabilitation systems.  
For example, (i) the excitation-inhibition property of feedback signals determines the sensitivity pattern while the activation-inactivation property determines the performance pattern; (ii) the strong nonlinearity of feedback activation with respect to biomechanical variables may contribute to achievable performance-sensitivity optima; (iii) force-dependent feedback can prevent the performance/robustness tradeoffs commonly occuring with the length-dependent feedback.  
These findings may yield important information for future work modeling biphasic rhythm generation, in that they provide insights that could guide the design of feedback systems to accomplish well-balanced efficient and robust powerstroke-recovery activities in biological and robotic experiments.

The rest of this paper is organized as follows.
We present the mathematical formulation of the control problem we consider, together with the variational analysis approach we provide, in Section \ref{sec:mathematical_formulation}.
The two neuromechanical models, their feedback architectures, and their performance-sensitivity patterns are discussed in Section \ref{sec:HCO model application} and Section \ref{sec:Markin model application}, respectively.
Finally we summarize the broad framework as well as the main observations and insights that we obtain from the models in Section \ref{sec:discussion}, where we also discuss limitations, connections to previous literature, and possible implications of our results for biology and engineering as well as future directions.

\section{Mathematical formulation}
\label{sec:mathematical_formulation}

The motor systems we consider integrate central neural circuitry, biomechanics, and sensory inputs from the periphery to form a closed-loop control system.  
The model systems we study fall within the following general framework \citep{shaw2015significance,lyttle2017robustness,yu2021dynamical}:
\begin{equation}
\label{eq:general_system}
    \frac{d\mba}{dt}=\mbf(\mba)+\mbg(\mba,\mbx),\qquad
    \frac{d\mbx}{dt}=\mbh(\mba,\mbx)+ \mbj(\mbx,\kappa).
\end{equation}
Here, $\mba$ and $\mbx$ are vectors representing the neural activity variables and mechanical state variables, respectively; the vector field $\mbf(\mba)$ represents the intrinsic neural dynamics of the central pattern generator when isolated from the rest of the body; $\mbh(\mba,\mbx)$ captures the biomechanical dynamics driven by the central inputs; $\mbg(\mba,\mbx)$ carries the sensory feedback from the periphery, which modulates the neural dynamics; $\mbj(\mbx,\kappa)$ is an externally applied load to the mechanical variables controlled by a load parameter $\kappa$.

In the powerstroke-recovery systems, we assume the load interacts with the mechanics only during the powerstroke phase.
The portion of the trajectory comprising the powerstroke phase is specified separately for each model. 
For the purposes of this paper we assume the vector fields $\mbf,\mbg,\mbh,\mbj$ are sufficiently smooth (e.g.~twice differentiable) except at a finite number of transition surfaces -- for example at points marking the powerstroke-recovery transitions.

For many naturally occurring control systems, the mechanical variables $\mbx$ may include both the position and velocity of different body components, as well as muscle activation variables.
The sensory feedback function $\mbg$ may be difficult to ascertain experimentally. 
For example, the feedback could have an excitatory effect on the neural dynamics, or an inhibitory effect, or a mixture at different points within a single movement; it may depend not only on neural outputs but also the length, velocity, or tension of the mechanical components; it may arise from multiple channels each with different gain.
Given the broad varieties of the possible feedback functions, we restrict the scope of our investigation to some biologically plausible forms for the specific models we consider in sections~\ref{sec:HCO model application} and \ref{sec:Markin model application}.

\subsection{Performance and sensitivity}

Suppose for $\kappa\in\mathcal{I}\subset\mathbb{R}$, system \eqref{eq:general_system} has an asymptotically stable limit cycle solution $\gamma_\kappa(t)$ with period $T_\kappa$.
Let $q_\kappa$ represent the rate at which the system advances relative to the outside world, and let $y_\kappa$ represent the total progress achieved over one limit cycle,  i.e.,
$$y_\kappa=\int_0^{T_\kappa} q_\kappa(\gamma_\kappa(t))\,dt.$$
We note that in general,  the instantaneous performance measure $q$ may depend both on the system variables $\mbx$ and on  the control parameter $\kappa$.  

We consider the task performance of the system, denoted by $Q$, to be the progress divided by the limit-cycle period, or equivalently, the mean value of the rate of progress averaged around the limit cycle, defined as follows
\begin{align}
\label{eq:performance_def_general}
    Q(\kappa)=\frac{1}{T_\kappa}\int_0^{T_\kappa} q_\kappa(\gamma_\kappa(t))\,dt=\frac{y_\kappa}{T_\kappa}.
\end{align}
For a powerstroke-recovery system, we choose the time coordinate so that $t=0$ coincides with the beginning of the powerstroke phase, and write $T_\kappa^\text{ps}$ for the duration of the power stroke.  
We adopt the convention that during the recovery phase, the position with respect to the outside world is held fixed ($q_\kappa\equiv 0$)\footnote{This convention is consistent with a simplified single-limb swing-stance model \citep{markin2010afferent,spardy2011dynamical} as well as with a simplified model of feeding biomechanics in \textit{Aplysia californica} \citep{shaw2015significance,lyttle2017robustness,wang2022variational}}, and denote $T_\kappa^\text{re}$ as the recovery phase duration  ($T_\kappa^\text{ps}+T_\kappa^\text{re}=T_\kappa$).
In such a system, we write the performance \eqref{eq:performance_def_general} as 
\begin{align}
\label{eq:performance_def}
    Q(\kappa)=\frac{1}{T_\kappa}\int_0^{T_\kappa^\text{ps}} q_\kappa(\gamma_\kappa(t))\,dt.
\end{align}

Assume that $\kappa_0\in\mathcal{I}$, which represents the unperturbed load.
When the system is subjected to a small static perturbation on the load, $\kappa_0\rightarrow\kappa_\epsilon=\kappa_0+\epsilon\in\mathcal{I}$, the original limit cycle trajectory $\gamma_{\kappa_0}$ is shifted to a new trajectory $\gamma_{\kappa_\epsilon}$, and its ability to resist the external change to maintain the performance is considered as a measure of robustness for the system.
Since the sensory feedback pathways regulate the system dynamics, it would be desirable to obtain a feedback function $\mbg$ so that the system is most robust against the load change, i.e.,
\begin{equation}
\label{eq:minimization_Q}
\mbg^*(\mba,\mbx)=\underset{\mbg(\mba,\mbx)}{\operatorname{argmin}}\,|Q(\kappa_\epsilon)-Q(\kappa_0)|.
\end{equation}
Suppose we can expand $Q(\kappa_\epsilon)$ around $\kappa_0$:
\begin{equation*}
    Q(\kappa_\epsilon)=Q(\kappa_0)+\epsilon\frac{\partial Q}{\partial \kappa}(\kappa_0)+O(\epsilon^2).
\end{equation*}
Then the minimization problem \eqref{eq:minimization_Q} to the first order reduces to
\begin{equation*}
\mbg^*(\mba,\mbx)=\underset{\mbg(\mba,\mbx)}{\operatorname{argmin}}\,\left|\epsilon\frac{\partial Q}{\partial \kappa}(\kappa_0)\right|=\underset{\mbg(\mba,\mbx)}{\operatorname{argmin}}\,\left|\frac{\partial Q}{\partial \kappa}(\kappa_0)\right|.
\end{equation*}
We quantify the sensitivity of the original system to be $S=\left|\frac{\partial Q}{\partial \kappa}(\kappa_0)\right|$, which describes the (infinitesimal) response of the task performance to the external perturbation on the load.
When $S=0$ with a certain feedback function, $Q(\kappa_\epsilon)\approx Q(\kappa_0)$, which implies a strong ability of the system to maintain performance homeostasis\footnote{\emph{Homeostasis} refers to the situation when a quantity remains approximately constant as a parameter varies over some range \citep{golubitsky2017homeostasis,golubitsky2018homeostasis,golubitsky2020infinitesimal}. For limit cycle systems, \cite{yu2022homeostasis} defined a homeostasis criterion in terms of the zero derivative(s) of the \emph{averaged} quantity with respect to the perturbation parameter.}. 
Define a (linear) functional $J: q\rightarrow S$, and the problem falls into a functional minimization problem, $\min_\mbg J[\mbg]$, which attains the minimum with some feedback function $\mbg^*$ when the functional derivative $\partial J/\partial\mbg^*=0$.  Finding $\mbg^*$ when the underlying system has a limit cycle (as opposed to the more often studied case of fixed-point homeostasis) is an open problem that we do not attempt to solve here.  
Instead, we focus on a limited range of $\mbg$ functions with practical significance and investigate the constrained optimization problem for the optimal feedback structure within that range.

\subsection{Variational analysis}

Variational analysis using the \emph{infinitesimal shape response curve} (iSRC) and the \emph{local timing response curve} (lTRC) is the key tool in our derivation of the performance sensitivity $S=\left|\frac{\partial Q}{\partial \kappa}(\kappa_0)\right|$.
In this section, we present a brief review of the theory and then provide two analytical methods to calculate the sensitivity.
More mathematical details are given in Appendix~\ref{app:variational_analysis}.

Consider system \eqref{eq:general_system} written in the form 
\begin{equation}
\frac{d\mbz}{dt}=\mbF_\kappa(\mbz),
\end{equation}
where $\mbz=(\mba,\mbx)^\intercal$ and $\mbF_\kappa(\mbz)$ is the corresonding vector field parameterized by the load $\kappa\in\mathcal{I}$.
For convenience, we write trajectories $\gamma_{\kappa_\epsilon}$ as $\gamma_{\epsilon}$ and $\gamma_{\kappa_0}$ as $\gamma_{0}$; we write other quantities similarly.
Expanding the perturbed trajectory yields
\begin{equation}
\label{eq:gamma_epsilon}
\gamma_\epsilon(\tau_\epsilon(t))=\gamma_0(t)+\epsilon\gamma_1(t)+O(\epsilon^2).
\end{equation}
The timing function $\tau_\epsilon(t)$ rescales the perturbed trajectory to match the unperturbed trajectory, so that the series \eqref{eq:gamma_epsilon} is uniform with respect to the time coordinate.
The linear shift in the shape of the unperturbed trajectory, $\gamma_1(t)$, is referred to as the iSRC. 
It satisfies a nonhomogeneous variational equation \citep{wang2021shape,yu2022homeostasis}
\begin{equation}
\label{eq:isrc_ode}
\frac{d\gamma_1(t)}{dt}=D\mbF_0(\gamma_0(t))\gamma_1(t)+\nu_1(t)\mbF_0(\gamma_0(t))+\frac{\partial\mbF_\kappa(\gamma_0(t))}{\partial\kappa}\bigg|_{\epsilon=0},
\end{equation}
where $\nu_1(t)=\frac{\partial^2\tau_\epsilon(t)}{\partial\epsilon\partial t}\big|_{\epsilon=0}$ measures the local timing sensitivity to the perturbation.
The initial condition for the iSRC equation \eqref{eq:isrc_ode} is 
\begin{equation}
\label{eq:isrc_initial}   \gamma_1(0)=\lim_{\epsilon\rightarrow0}\frac{\mbp_\epsilon-\mbp_0}{\epsilon},
\end{equation}
where $\mbp_\epsilon$ and $\mbp_0$ represent the intersection points of the trajectories with an arbitrary Poincar\'e section transverse to both the perturbed and unperturbed limit cycles, so that $\gamma_1(0)$ indicates the linear displacement of the unperturbed intersection point.
For more details about the iSRC, see Appendix~\ref{subapp:iSRC}, \cite{wang2021shape}, and \cite{yu2022homeostasis}.

To solve the iSRC equation \eqref{eq:isrc_ode}, the lTRC is built to yield the timing sensitivity $\nu_1$ local to each phase of the motion.
Approximate the perturbed phase durations by
\begin{align*}
    T_\epsilon^\text{ps}&=T_0^\text{ps}+\epsilon T_1^\text{ps}+O(\epsilon^2),\\
    T_\epsilon^\text{re}&=T_0^\text{re}+\epsilon T_1^\text{re}+O(\epsilon^2).
\end{align*}
\cite{wang2021shape} and \cite{yu2023sensitivity} developed a formula to calculate the first-order approximation for the duration change in phase $i\in\{\text{ps},\text{re}\}$, given by
\begin{equation}
\label{eq:T1_formula}
    T_1^i=\eta^i(\mbz_0^\text{in})\cdot\frac{\partial\mbz_\kappa^\text{in}}{\partial\kappa}\bigg|_{\epsilon=0}-\eta^i(\mbz_0^\text{out})\cdot\frac{\partial\mbz_\kappa^\text{out}}{\partial\kappa}\bigg|_{\epsilon=0}+\int_{t^\text{in}}^{t^\text{out}}\eta^i(\gamma_0(t))\cdot\frac{\partial\mbF_\kappa(\gamma_0(t))}{\partial\kappa}\bigg|_{\epsilon=0}\,dt.
\end{equation}
Here, $\mbz_0^\text{in}$ and $\mbz_0^\text{out}$ denote the unperturbed entry point to, and exit point from, the specific phase, respectively. 
Vector $\eta^i$, defined to be the gradient of the remaining time of the trajectory until exiting phase $i$, is referred to as the lTRC for phase $i$.
It satisfies the adjoint equation
\begin{equation*}
    \frac{d\eta^i}{dt}=-D\mbF_0(\gamma_0(t))^\intercal\eta^i,
\end{equation*}
with a boundary condition
$$\eta^i(\mbz_0^\text{out})=-\frac{n^\text{out}}{(n^\text{out})^\intercal\mbF_0(\mbz_0^\text{out})},$$
where $n^\text{out}$ is a normal vector of the exit boundary surface at $\mbz_0^\text{out}$.
When the vector field $\mbF$ changes discontinuously across the surface defining the boundary between two regions, the Jacobian $D\mbF$ should be evaluated as a one-sided limit, taken from the interior of the local region.
With a linear time scaling for \eqref{eq:gamma_epsilon} (and setting $t=0$ to be the start of the power stroke), i.e., 
\begin{align*}
&\tau_\epsilon^\text{ps}(t)=\frac{T_\epsilon^\text{ps}}{T_0^\text{ps}}t,\qquad t\in[0,T_0^\text{ps}),\\
&\tau_\epsilon^\text{re}(t)=\tau_\epsilon^\text{ps}(T_0^\text{ps})+\frac{T_\epsilon^\text{re}}{T_0^\text{re}}(t-T_0^\text{ps}),\qquad t\in[T_0^\text{ps},T_0),
\end{align*}
the local timing sensitivity function in \eqref{eq:isrc_ode} reduces to
$$\nu_1^\text{ps}=\frac{T_1^\text{ps}}{T_0^\text{ps}},\qquad\nu_1^\text{re}=\frac{T_1^\text{re}}{T_0^\text{re}},$$
which can be obtained by using equation \eqref{eq:T1_formula} for each phase.
See Appendix~\ref{subapp:lTRC}, \cite{wang2021shape}, and \cite{yu2023sensitivity} for more details about the lTRC formulation.

The variational analysis above allows us to analyze the performance sensitivity of 
powerstroke-recovery systems.
In \cite{yu2022homeostasis} we provided a formula for the sensitivity of any averaged quantity with respect to an arbitrary control parameter, as long as the quantity of interest does not have explicit dependence on the control parameter.  
We generalize the approach of \cite{yu2022homeostasis} to allow for dependence of the instantaneous performance $q_\kappa(\mbx)$ on both the state $\mbx$ and the parameter $\kappa$, and obtain
\begin{equation}
\label{eq:sensitivity_formula1}
    \frac{\partial Q}{\partial\kappa}(\kappa_0)=\frac{1}{T_0^\text{ps}}\int_0^{T_0^\text{ps}}\left[\beta_0\left(\nabla q_{0}(\gamma_0(t))\cdot\gamma_1(t)+\frac{\partial q_\kappa(\gamma_0(t))}{\partial\kappa}\bigg|_{\epsilon=0}\right)+\beta_1q_0(\gamma_0(t))\right]\,dt.
\end{equation}
The first term in the integral arises from the impact of the perturbation on the shape of the trajectory ($\gamma_1$) as well as directly on the quantity of interest ($\partial q_\kappa/\partial \kappa)$.
Here $\beta_0$ denotes the proportion of the powerstroke duration within the period ($\beta_0=T_0^\text{ps}/T_0$).
The second term indicates the impact of the perturbation on the timing of the trajectory, in that $\beta_1$ represents the linear shift in $\beta_0$ in response to the perturbation, which can be analytically evaluated by
\begin{equation*}
\label{eq:beta_1}
\beta_1=\frac{\partial\beta_\kappa}{\partial\kappa}\bigg|_{\epsilon=0}=\frac{T_1^\text{ps}T_0-T_0^\text{ps}T_1}{T_0^2}.
\end{equation*}
The derivation of formula \eqref{eq:sensitivity_formula1} is given in Appendix~\ref{subapp:sensitivity_formula1}.

Given the special structure of powerstroke-recovery systems, we can derive a more succinct expression for $\partial Q/\partial \kappa$.  
For any value of $\kappa,$ the 
second definition in \eqref{eq:performance_def_general} gives 
\begin{align}
\label{eq:derivative_Q}
    \frac{\partial Q}{\partial\kappa}(\kappa)=\frac{1}{T_\kappa^2}\left(\frac{\partial y_\kappa}{\partial\kappa}T_\kappa-y_\kappa\frac{\partial T_\kappa}{\partial\kappa}\right)=\frac{y_\kappa}{T_\kappa}\left(\frac{1}{y_\kappa}\frac{\partial y_\kappa}{\partial\kappa}-\frac{1}{T_\kappa}\frac{\partial T_\kappa}{\partial\kappa}\right)=Q(\kappa)\left(\frac{1}{y_\kappa}\frac{\partial y_\kappa}{\partial\kappa}-\frac{1}{T_\kappa}\frac{\partial T_\kappa}{\partial\kappa}\right).
\end{align}
Recall at $\kappa=\kappa_\epsilon$, we write $y_\epsilon$ and $T_\epsilon$ as shorthand for $y_{\kappa_\epsilon}$ and $T_{\kappa_\epsilon}$.
We can expand the perturbed progress $y_\epsilon$ and period $T_\epsilon$ around $\epsilon=0$ as
\begin{align*}
    y_\epsilon&=y_0+\epsilon y_1+O(\epsilon^2),\\
    T_\epsilon&=T_0+\epsilon T_1+O(\epsilon^2),
\end{align*}
where $y_1$ is approximately given by the net change of the mechanical component of the iSRC $\gamma_1$ (cf.~equations~\eqref{eq:isrc_ode} and \eqref{eq:isrc_initial}) within the powerstroke phase, and $T_1$ is the linear shift in the total period, readily given by $T_1=T_1^\text{ps}+T_1^\text{re}$ (cf.~equation~\eqref{eq:T1_formula}).
Therefore, equation~\eqref{eq:derivative_Q} at $\kappa=\kappa_0$ becomes
\begin{align}
\label{eq:sensitivity_formula2}
    \frac{\partial Q}{\partial\kappa}(\kappa_0)=Q_0\left(\frac{y_1}{y_0}-\frac{T_1}{T_0}\right),
\end{align}
which, like \eqref{eq:sensitivity_formula1}, incorporates both the shape and timing effects of the perturbation in two distinct terms.
Equation~\eqref{eq:sensitivity_formula2} also suggests that the sensitivity can be directly given by the absolute difference between the first-order timing change and shape change induced by the perturbation.
When the two effects completely offset each other, the system achieves ``perfect" robustness.

The two expressions given by \eqref{eq:sensitivity_formula1} and \eqref{eq:sensitivity_formula2}  for calculating the sensitivity of the task performance for powerstroke-recovery systems allow us to compare different sensory feedback mechanisms in pursuit of an efficient and robust motor pattern.
In the following sections we develop two illustrative examples: an abstract CPG-motor model introduced in \cite{yu2021dynamical}, and an unrelated realistic locomotor model studied in \cite{markin2010afferent} and \cite{spardy2011dynamical}.
The two examples show a variety of differences in their model construction, but our analytic framework is broad enough to address both and give useful insights.
Simulation codes required to produce each figure are available at\\ \texttt{https://github.com/zhuojunyu-appliedmath/Powerstroke-recovery}.

\section{Application: HCO model with  external load}
\label{sec:HCO model application}

\begin{figure}
\centering
\includegraphics[width=14cm]{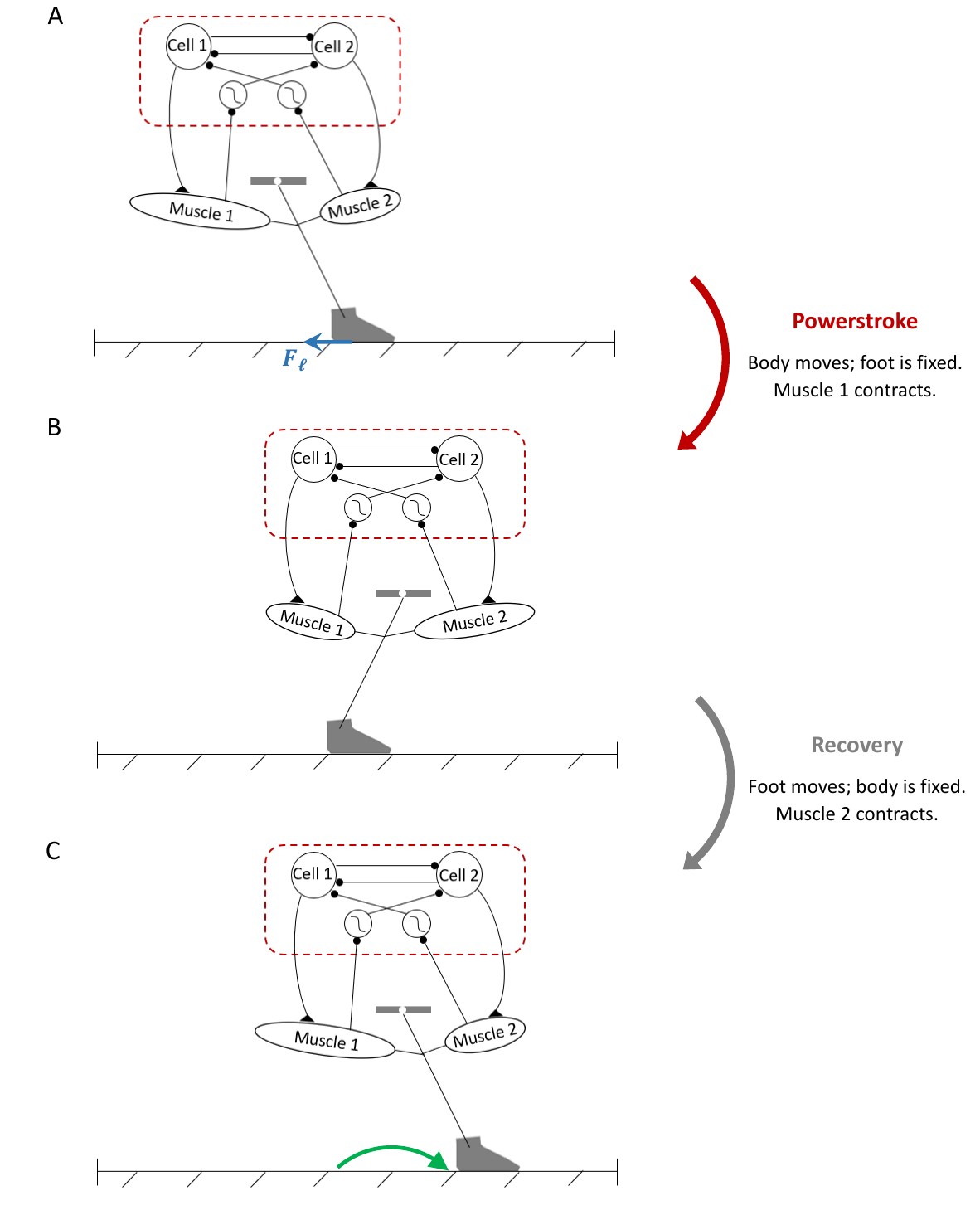}
\caption{\label{fig:Yu_sketch} Schematic of components and basic behavior of the HCO model system, adapted from \cite{yu2021dynamical}.
The CPG circuit of the system comprises a half-center oscillator, represented by mutually inhibitory cell 1 and cell 2.
Output from each neuron drives its ipsilateral muscle pulling a limb.
The muscle stretch and contraction in turn produce reflex commands to a feedback receptor, which sends an inhibitory signal to its contralateral neuron. 
The limb interacts with an external substrate, which imposes a mechanical load opposing the limb movement.
Inhibitory connections end with a round ball, and excitatory connections end with a triangle.
In a single movement cycle, the powerstroke phase (panel A to panel B) occurs when the body (red dashed rectangle) moves forward while the foot is fixed, subjected to the load $F_\ell$ opposite to the movement direction (blue arrow).
The recovery phase follows (panel B to panel C), during which the body is fixed and the foot is lifted off the substrate and repositions for the next power stroke (green arrow).} 
\end{figure}

In \cite{yu2021dynamical}, we studied a simple closed-loop model combining neural dynamics and biomechanics, as sketched in Fig.~\ref{fig:Yu_sketch}.
The CPG sytem comprises a half-center oscillator (HCO) with two conductance-based Morris–Lecar neurons \citep{morris1981voltage,skinner1994mechanisms}.
Outputs from the HCO drive a simple biomechanical system, which follows a Hill-type kinetic model based on experimental data from the marine mollusk \textit{Aplysia californica} \citep{yu1999biomechanical}.
Sensory feedback from the periphery couples the body and brain dynamics, allowing the system to interact with the changing outside world, and to modulate the central neural activities adaptively.
However, the previous study of this model did not explore the performance with respect to a physical task; rather, the CPG followed an autonomous clocklike pattern.
To perform a more meaningful analysis for understanding principles of closed-loop motor control, here we augment the model from \cite{yu2021dynamical} by incorporating a mechanical load exerted in a specific direction with recurrent engagement and disengagement with the system, which enables us to apply our quantitative measures of progress and sensitivity of the system.

\subsection{The equations of the HCO model}

The model equations we consider are as follows. For $i,j=1,2$ and $j\neq i$,
\begin{equation}
\label{eq:Yu_model_equations}
\begin{split}
C\frac{dV_i}{dt}&=I_\text{ext}-g_\text{L}(V_i-E_\text{L})-g_\text{Ca}M_\infty(V_i)(V_i-E_\text{Ca})-g_\text{K}N_i(V_i-E_\text{K})\\
&\qquad\quad-g_\text{syn}^\text{CPG}S_\infty^\text{CPG}(V_j)\left(V_i-E_\text{syn}^\text{CPG}\right)-g_\text{syn}^\text{FB}S_\infty^\text{FB}(L_j)(V_i-E_\text{syn}^\text{FB}),\\
\frac{dN_i}{dt}&=\lambda_N(V_i)(N_\infty(V_i)-N_i),\\
\frac{dA_i}{dt}&=\tau^{-1}\{U(V_i)-[\beta+(1-\beta)U(V_i)]A_i\},\\
\frac{dx}{dt}&=\frac{1}{b}(F_2-F_1+r\kappa F_\ell).
\end{split}
\end{equation}
Variable $V_i$ denotes the membrane voltage for HCO neuron cell $i$, and $N_i$ is the gating variable for the potassium current in cell $i$. 
The two neuron cells are coupled by fast inhibitory synapses, and the coupling function is given by 
$$S_\infty^\text{CPG}(V_j)=\frac{1}{2}\left(1+\tanh\left(\frac{V_j-E_\text{thresh}}{E_\text{slope}}\right)\right),$$
which closely approximates a Heaviside step function with $E_\text{thresh}$ denoting the synaptic threshold.

In the third equation of \eqref{eq:Yu_model_equations}, $A_i\ge 0$ represents the  activation of the $i$th muscle.
The neural outputs from the HCO drive the associated muscle, modeled as
$$U(V_i)=1.03-4.31\exp{(-0.198(V_i/2))},\quad V_i\geq16.$$
The biomechanics is represented by the movement of an object (nominally, a pendulum or limb), with each side connected to one of the two muscles.
The object position relative to the center of mass of the organism, denoted as $x$, is controlled by the muscle forces $F_1$ and $F_2$ acting on it.
An external load $F_\ell$ is exerted on the object only during the powerstroke phase, as specified by the indicator variable $r$ defined to be
\begin{align*}
\label{eq:indicator_r}
r=\left\{\begin{aligned}
&1,\quad&\text{power stroke},\\
&0,\quad&\text{recovery}.
\end{aligned}\right.
\end{align*}
We assume that the powerstroke phase is at work when $V_1$ is in the active state ($V_1> E_\text{thresh}$), whereas the load is absent from the system when $V_1$ is inhibited ($V_1\leq E_\text{thresh}$).
Parameter $\kappa$ describes the strength of the load, which is considered as the perturbation parameter for this model.

\begin{figure}[htb]
\centering
\includegraphics[width=14cm]{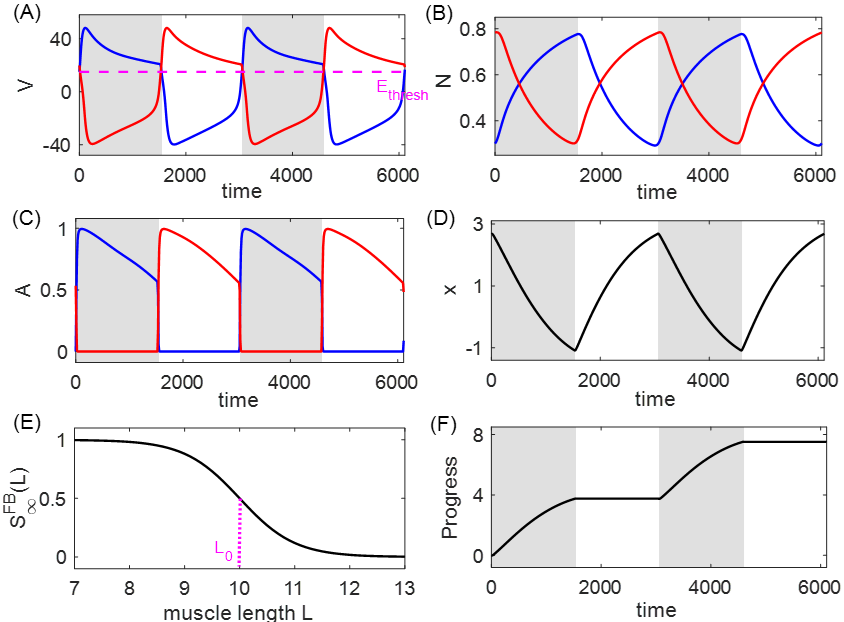}
\caption{\label{fig:Yu_sol_example} A typical solution for the system \eqref{eq:Yu_model_equations} in the absence of perturbation ($\kappa_0=1$), plotted over two periods with $T_0=3055$ ms. 
The system is governed by the inhibitory-contralateral feedback mechanism, with a decreasing sigmoid activation function $S_\infty^\text{FB}$ given by \eqref{eq:contra-dec-FB}, and threshold $L_0=10$ (panel E). 
Blue trace: cell/muscle 1. Red trace: cell/muscle 2. 
The gray shaded regions represent the powerstroke phase, defined by the active state of cell 1 ($V_1>E_\text{thresh}$, panel A), with duration $T_0^\text{ps}=1544$ ms; the white regions represent the recovery phase with duration $T_0^\text{re}=1511$ ms.
The system makes progress during the powerstroke phase at rate $q=-dx/dt$, while it maintains its position during the recovery phase (compare panels D and F).} 
\end{figure}

The system completes an intact closed loop through the sensory feedback induced by the biomechanics on the CPG in the form of feedback currents, e.g.,
$$g_\text{syn}^\text{FB}S_\infty^\text{FB}(L_j)(V_i-E_\text{syn}^\text{FB}),$$
where $L_j$ is the length of muscle $j$, and the
function $S_\infty^\text{FB}(L_j)$ describes the feedback synaptic activation.
We assume that the feedback conductance has fast dynamics, following a sigmoid function.
As discussed in \cite{yu2021dynamical}, the feedback synaptic architecture affords eight variations, depending on whether the feedback is (i) inhibitory or excitatory, (ii) activated by muscle contraction or muscle stretch, and (iii) modulatory on the contralateral or ipsilateral neuron.
For example, when the feedback current is inhibitory to its contralateral neuron and activated when the muscle is contracted, then for the $V_i$-equation we set $E_\text{syn}^\text{FB}=-80$ mV and
\begin{equation}
\label{eq:contra-dec-FB}
    S_\infty^\text{FB}(L_j)=\frac{1}{2}\left(1-\tanh\left(\frac{L_j-L_0}{L_\text{slope}}\right)\right).
\end{equation}
Figure~\ref{fig:Yu_sketch} illustrates the system controlled by the inhibitory-contralateral-decreasing feedback mechanism, and Fig.~\ref{fig:Yu_sol_example} shows a typical solution for the system.
In contrast, setting $E_\text{syn}^\text{FB}=80$ mV for the excitatory feedback current, or setting the sigmoid function to be increasing for the muscle-stretch activated case, or changing $L_j$-dependence to $L_i$-dependence for the ipsilateral mechanism, would specify other possible mechanisms.
We compare the performance and sensitivity of all different realizations of each of the eight variations of the feedback control scheme below.
The force terms $F_{1,2}$, as well as additional details about the functions in system~\eqref{eq:Yu_model_equations}, parameter values used for simulations, and simulation codes are given in Appendix~\ref{app:Yu_model_details}.

\subsection{Analysis of the HCO model}

\begin{figure}
\centering
\includegraphics[width=14cm]{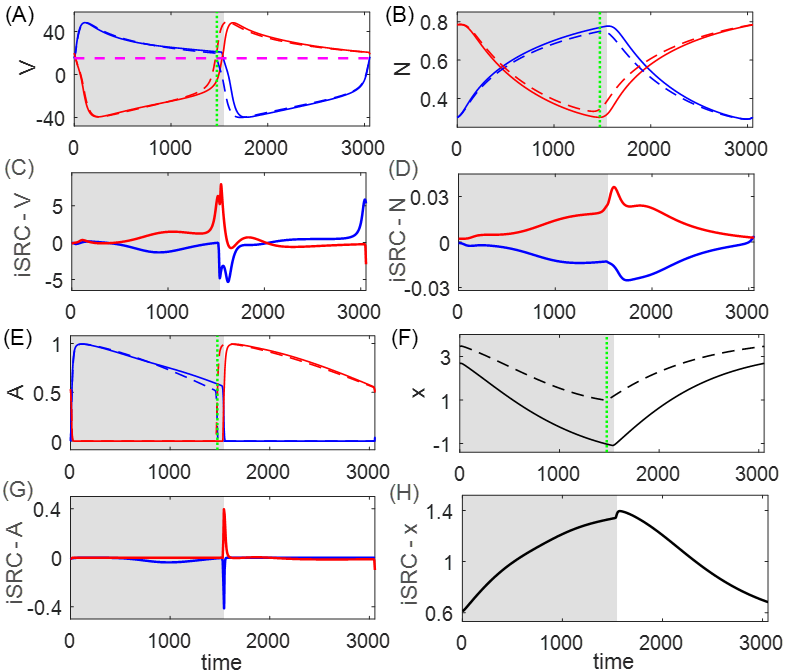}
\caption{\label{fig:Yu_iSRC_example} The iSRC analysis for system \eqref{eq:Yu_model_equations}.
\textbf{A, B, E, F}: Time series of the unperturbed trajectory (solid, same as Fig.~\ref{fig:Yu_sol_example}) and perturbed trajectory with load $\kappa_\epsilon=2$ (dashed), both of which are initiated at the start of their respective powerstroke phase.
The perturbed trajectory is uniformly time-rescaled by $t_\epsilon=\frac{T_0}{T_\epsilon}t$ to compare with the unperturbed trajectory, where $T_0=3055$ and $T_\epsilon=2831$.
The shaded regions represent the powerstroke phase for the unperturbed case, whereas the vertical green dotted lines represent the time at which the phase for the perturbed case switches from power stroke to recovery, which is advanced due to the perturbation. 
\textbf{C, D, G, H}: The components for the iSRC $\gamma_1(t)$ of the unperturbed trajectory, with the initial condition defined by the Poincar\'e section $\{V_1=0,\,dV_1/dt>0\}$. 
The negative responses in the $V_1, N_1, A_1$ directions at the end of power stroke are consistent with the earlier  transition to the recovery phase shown in panels A, B, C.
The accumulating positive response of $x$ results from the direct effect of perturbation, which shows a stronger resistance to the limb movement.} 
\end{figure}

In order to establish measures of performance and sensitivity, we assume that the system advances only during the powerstroke phase.  
That is, the rate at which the system makes progress is given by
\begin{equation*}
q=-r\frac{dx}{dt},
\end{equation*}
as indicated in Fig.~\ref{fig:Yu_sol_example}D, F.
Therefore, by~\eqref{eq:performance_def}, the performance (i.e., the average rate of progress) is 
\begin{align}
\label{eq:performance_Yu}
    Q(\kappa)=\frac{1}{T_\kappa}\int_0^{T_\kappa}\left(-r\frac{dx}{dt}\right)\,dt=-\frac{1}{T_\kappa b\ell}\int_0^{T_\kappa^\text{ps}}(F_2-F_1+\kappa F_\ell)\,dt.
\end{align}
When a sustained small perturbation is applied to the load, $\kappa_0\rightarrow\kappa_0+\epsilon$, the solution trajectory shifts in both its shape and timing, and the performance of the perturbed system is consequently different from the performance of the unperturbed system.
As a reference, Fig.~\ref{fig:Yu_iSRC_example}A, B, E, F compare the trajectories of the unperturbed solution with $\kappa_0=1$ for the system shown in Fig.~\ref{fig:Yu_sol_example} and the perturbed solution with $\kappa_\epsilon=2$, and Fig.~\ref{fig:Yu_iSRC_example}C, D, G, H illustrate the iSRC $\gamma_1$ of the unperturbed trajectory, specified by Poincar\'e section $\{V_1=0,\,dV_1/dt>0\}$.
Note that for visual convenience, the large perturbation ($\epsilon=1$) is applied here, but in our actual analysis the perturbation magnitude should be small ($|\epsilon|\ll1$).
Our analysis yields several observations about the role of the inhibition-contralateral-decreasing sensory feedback in regulating the system's response to the perturbation, as discussed in detail below.

With the perturbation (increased load), the transition from the power stroke to recovery is advanced, i.e., the powerstroke phase is shorter.
As indicated in Fig.~\ref{fig:Yu_iSRC_example}F, the immediate effect of the perturbation is the positive displacement and slower change rate in the $x$-variable, which occurs because the object is being pulled by the stronger load opposite to its movement direction.
Correspondingly, muscle 2, whose length is $L_2=10-x$, is more contracted than it is in the unperturbed case. 
The sensory feedback current injected to neuron 1, with synaptic activation given by \eqref{eq:contra-dec-FB}, is therefore larger and gives more inhibition to the active neuron 1.
As a result, the active $V_1$ crosses the synaptic threshold and terminates the powerstroke phase at an earlier time, as indicated in Fig.~\ref{fig:Yu_iSRC_example}A.
The iSRC in each direction is consistent with the associated trajectory comparison.
The significant negative peak in the $V_1$-component at the end of powerstroke phase (panel C) suggests that $V_1$ of the perturbed solution at the rescaled time already decreases to the synaptic threshold and jumps down to the inhibited state, indicating the transition out of the power stroke is advanced.
Since $\kappa$ directly impacts $dx/dt$, we see a different effect on  $x$ (panel H) than the other variables. 

\begin{figure}
\centering
\includegraphics[width=16.5cm]{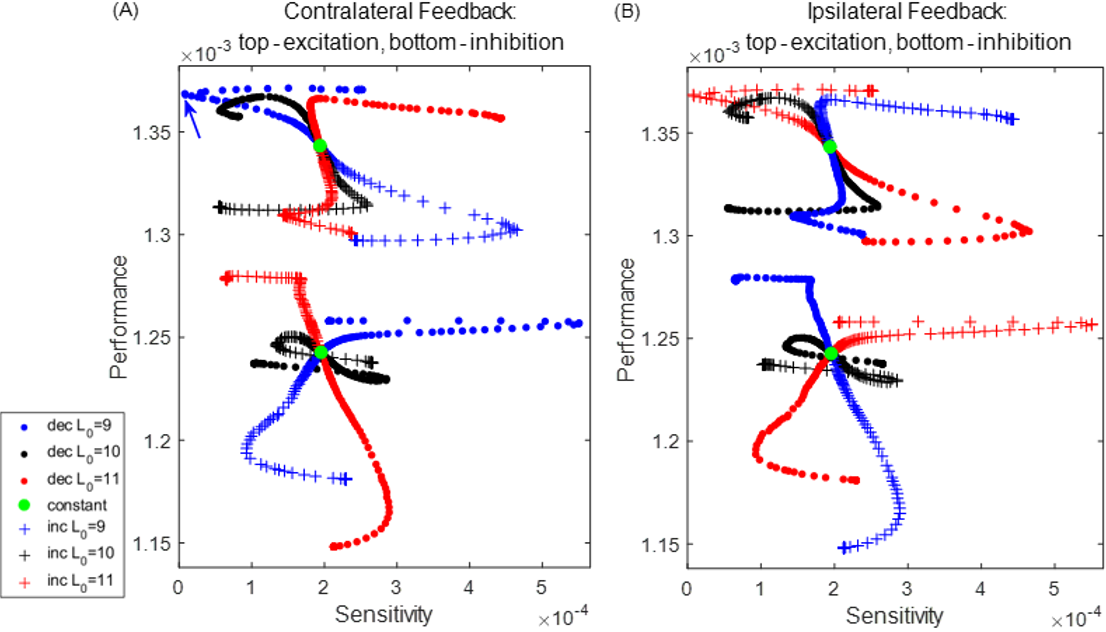}
\caption{\label{fig:Yu_per_sen} Performance-sensitivity patterns of all eight sensory feedback architectures in the HCO model.
(A) Contralateral feedback. (B) Ipsilateral feedback. 
Each panel includes four subarchitectures: excitatory feedback (upper ensembles) \textit{versus} inhibitory feedback (lower ensembles), and inactivating feedback (dots) \textit{versus} activating feedback (+ signs).
Specifically, the dot trace indicates decreasing synaptic feedback activation $S_\infty^\text{FB}$ (muscle-contraction activated feedback), while the plus trace indicates increasing activation  $S_\infty^\text{FB}$ (muscle-stretch activated feedback).
The half-threshold position parameter $L_0$ of $S_\infty^\text{FB}$ is varied over $\{9, 10, 11\}$, colored by blue, black, and red, respectively. 
A green dot at the center of each ensemble indicates the system with constant feedback ($S_\infty^\text{FB}\equiv 0.5$). 
Starting from this constant feedback case, the performance-sensitivity curve of each mechanism becomes distinct as the shape of $S_\infty^\text{FB}$ becomes steeper (i.e., $L_\text{slope}$ decreases) until it approaches a Heaviside step function ($L_\text{slope}\rightarrow0$).
The contralateral feedback mechanism with  increasing (resp. decreasing) sigmoidal activation $S_\infty^\text{FB}$ with $L_0=10+\theta$ is functionally equivalent to the ipsilateral mechanism with decreasing (resp. increasing) sigmoidal activation $S_\infty^\text{FB}$ with $L_0=10-\theta$, reducing the eight feedback architectures to four fundamentally different classes. 
Blue arrow (panel A top) marks the most advantageous configuration among those tested.} 
\end{figure}

Both efficiency (high performance) and robustness (low sensitivity) are important features of motor control systems interacting with the outside world.
The performance for the perturbed system in Fig.~\ref{fig:Yu_iSRC_example} is smaller than the unperturbed system ($Q_\epsilon=0.87\times10^{-3}$, $Q_0=1.23\times10^{-3}$), due to the larger magnitude in the progress decrease relative to that in the period.
To measure the ability of maintaining the performance, we quantify the sensitivity of the original system in response to an infinitesimal sustained perturbation, following \eqref{eq:sensitivity_formula1}, to be
\begin{equation}
\label{eq:sensitivity_Yu}
\frac{\partial Q}{\partial\kappa}(\kappa_0)=\frac{1}{T_0^\text{ps}}\int_0^{T_0^\text{ps}}\left[-\beta_0\left(\nabla \left(\frac{dx}{dt}\right)\cdot\gamma_1(t)+\frac{F_\ell}{b\ell}\right)-\beta_1\left(\frac{dx}{dt}\right)\right]\,dt.
\end{equation}
One can also estimate the sensitivity by \eqref{eq:sensitivity_formula2}, where $y_1$ corresponds to the $x$-component of $\gamma_1$. 
That is,
\begin{align*}
    \frac{\partial Q}{\partial\kappa}(\kappa_0)=Q_0\left(\frac{\gamma_{1,x}}{\int_0^{T_0^\text{ps}}(dx/dt)\,dt}-\frac{T_1}{T_0}\right).
\end{align*}
To evaluate the joint goals of high performance and low sensitivity, and to investigate how they are affected by  sensory feedback, we will simultaneously study the two measures plotted together, while manipulating the shape of the feedback activation function.
Specifically, we will vary the steepness parameter $L_\text{slope}$ and position/half-threshold parameter $L_0$ of the sigmoid synaptic feedback activation function $S_\infty^\text{FB}$.
Figure~\ref{fig:Yu_per_sen} shows the results for all eight sensory feedback mechanisms.

The eight superficially distinct feedback architectures can be reduced to four fundamentally different mechanisms in terms of their performance and sensitivity.
The contralateral mechanism of muscle-stretch activated (increasing) current with threshold $L_0=10+\theta$ ($\theta\in\mathbb{R}$), is equivalent to the ipsilateral mechanism of muscle-contraction activated (decreasing) current with threshold $L_0=10-\theta$. Specifically, substituting the contralateral-increasing feedback activation with $L_0=10+\theta$ to the $V_1$-equation of system \eqref{eq:Yu_model_equations} yields
\begin{align*}
\frac{dV_1}{dt}&=\cdots-g_\text{syn}^\text{FB}S_\infty^\text{FB}(L_2)(V_1-E_\text{syn}^\text{FB})\\
&=\cdots-\frac{g_\text{syn}^\text{FB}}{2}\left(1+\tanh{\left(\frac{10-x-(10+\theta)}{L_\text{slope}}\right)}\right)(V_1-E_\text{syn}^\text{FB})\\
&=\cdots-\frac{g_\text{syn}^\text{FB}}{2}\left(1+\tanh{\left(\frac{-x-\theta}{L_\text{slope}}\right)}\right)(V_1-E_\text{syn}^\text{FB})\\
&=\cdots-\frac{g_\text{syn}^\text{FB}}{2}\left(1-\tanh{\left(\frac{x+\theta}{L_\text{slope}}\right)}\right)(V_1-E_\text{syn}^\text{FB})\\
&=\cdots-\frac{g_\text{syn}^\text{FB}}{2}\left(1-\tanh{\left(\frac{10+x-(10-\theta)}{L_\text{slope}}\right)}\right)(V_1-E_\text{syn}^\text{FB}),
\end{align*}
where the last equation is exactly the case for the ipsilateral-decreasing feedback with $L_0=10-\theta$.
Note that for the unloaded model in \cite{yu2021dynamical}, only the inhibition-excitation property of the feedback makes a fundamental difference regarding the stability and robustness of the system. 
However, when we incorporate mechanical interactions with an external substrate, the activating property of the feedback must be taken into account.
In the following, we will discuss the performance and sensitivity for the four contralateral feedback mechanisms, which we call \emph{inhibition-increasing} (II), \emph{inhibition-decreasing} (ID), \emph{excitation-increasing} (EI), and \emph{excitation-decreasing} (ED).
The other four ipsilateral feedback mechanisms can be applied accordingly.

\subsection{Performance and sensitivity of the HCO model}

Our analysis of the HCO model, subject to an applied external load, leads to the following observations from Fig.~\ref{fig:Yu_per_sen}: 
\begin{enumerate}
    \item Excitatory feedback is advantageous over inhibitory feedback in terms of performance.
    \item The qualitative patterns of performance are reversed with respect to the steepness of synaptic activation function in the activating and inactivating feedback mechanisms.
    \item The qualitative patterns of sensitivity are reversed with respect to the steepness of synaptic activation function in the excitatory and inhibitory feedback mechanisms.   
    \item When the sigmoid activation function $S_\infty^\text{FB}$ is approximately linear over the working range of the limb, the performance-sensitivity changes approximately linearly with the slope of $S_\infty^\text{FB}$. 
    \item As the working range of the limb extends beyond the linear regime of $S_\infty^\text{FB}$, the performance-sensitivity curve can become strongly nonlinear and even non-monotonic, leading to well-defined simultaneous optima in both performance and sensitivity.
\end{enumerate}
We discuss each of these points in turn below.

\begin{figure}
\centering
\includegraphics[width=14cm]{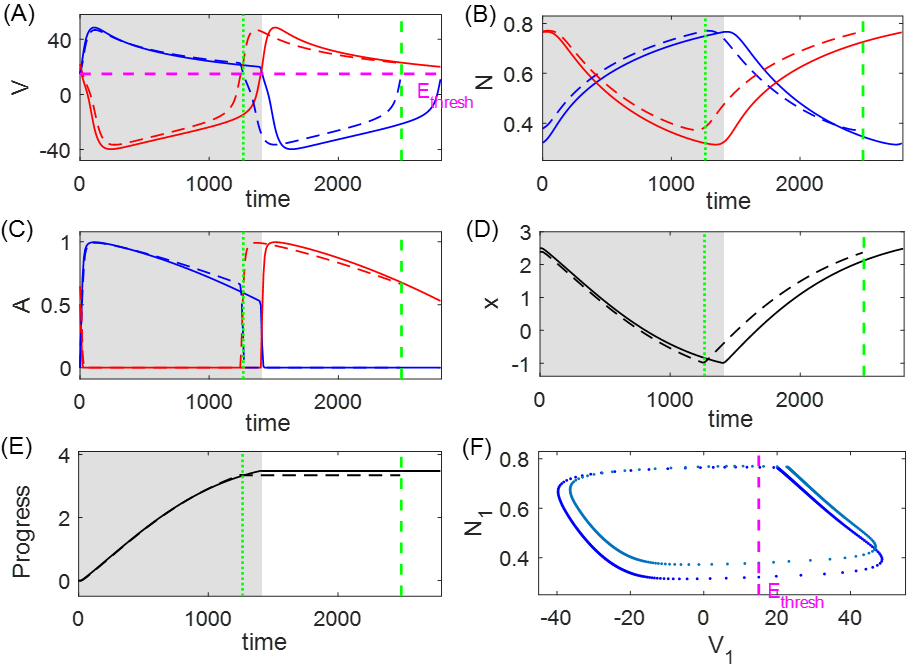}
\caption{\label{fig:Yu_ex_in_per} Comparison of trajectories for the constant inhibitory feedback system (solid lines) and constant excitatory feedback system (dashed lines),  plotted over one period.
The only difference in the setting for the two systems is the synaptic feedback reversal potential, $E_\text{syn}^\text{FB}=+80$ or $-80$ mV, respectively.
The gray shaded region represents the powerstroke phase of the inhibitory system; the vertical green dotted line denotes the transition time of the excitatory system out of the power stroke, and the vertical green dashed line denotes the end of its recovery phase. 
Panel F shows the projection of the trajectories to the $(V_1, N_1)$ phase plane, where the successive dots are equally spaced in time.
The inhibitory trajectory is in dark blue (generally passing through more negative voltages) and the excitatory trajectory in light blue (generally higher voltages).
The excitatory feedback current raises the voltage of the active neuron and greatly shortens both the powerstroke and recovery phases (panels A, F).
The limb makes smaller progress due to the shorter powerstroke duration (panels D, E), but the decrease in magnitude is smaller than the decrease in the period. 
Hence, the performance of the excitatory-feedback case is larger than the inhibitory-feedback case.
} 
\end{figure}

\paragraph{Excitatory sensory feedback outperforms inhibitory feedback.}
This conclusion is evident from Fig.~\ref{fig:Yu_per_sen}, in the higher location of the performance-sensitivity pattern for each of the excitation mechanisms relative to that for all of the inhibition mechanisms.
To understand the advantage of excitatory over inhibitory feedback in this model system, Fig.~\ref{fig:Yu_ex_in_per} compares the trajectories of two systems, one with excitatory constant feedback and the other with inhibitory constant feedback.
These systems correspond respectively to the two green dots in Fig.~\ref{fig:Yu_per_sen}A.
In the excitatory system, the extra excitation to $V_i$ due to the feedback has two opposing effects.
On the one hand, it advances the time at which the active neuron ``jumps down" to the inhibited state and thus shortens both the powerstorke phase and the total period $T$, relative to the system with constant inhibitory feedback.
Panel F of Fig.~\ref{fig:Yu_ex_in_per}  shows the projection of the trajectory on the $(V_1, N_1)$ plane with points plotted at constant time intervals.  
Note the rapid change in the voltage component relative to the slower gating variable.  
The significant timing change results from the exponential deceleration of the dynamics of the active neuron when approaching the jump-down point, as shown by the contraction of points before jumping in panel F.
Although the $(V_1, N_1)$ projection of the two trajectories during the active state does not differ much spatially, the difference in time needed to cover the small spatial difference is significant.
On the other hand, the constant excitatory drive also reduces the net progress $y$ of the system per cycle.
In particular, the progress $y$ of the system  declines due to the shorter movement time (panel E). 
The net performance ($Q=y/T$, equation~\eqref{eq:performance_def_general}) is determined by both the timing $T$ and shape $y$ of the trajectory.  
For the excitatory system, the resulting performance is in fact larger than for the inhibitory system, because the relative change in period is larger than the relative decrease in progress.
Therefore, regardless of the structure of the feedback pathway, any system equipped with the excitatory sensory feedback is always advantageous over the system with the inhibitory sensory feedback in terms of their performance.

\begin{figure}[htb]
\centering
\includegraphics[width=14cm]{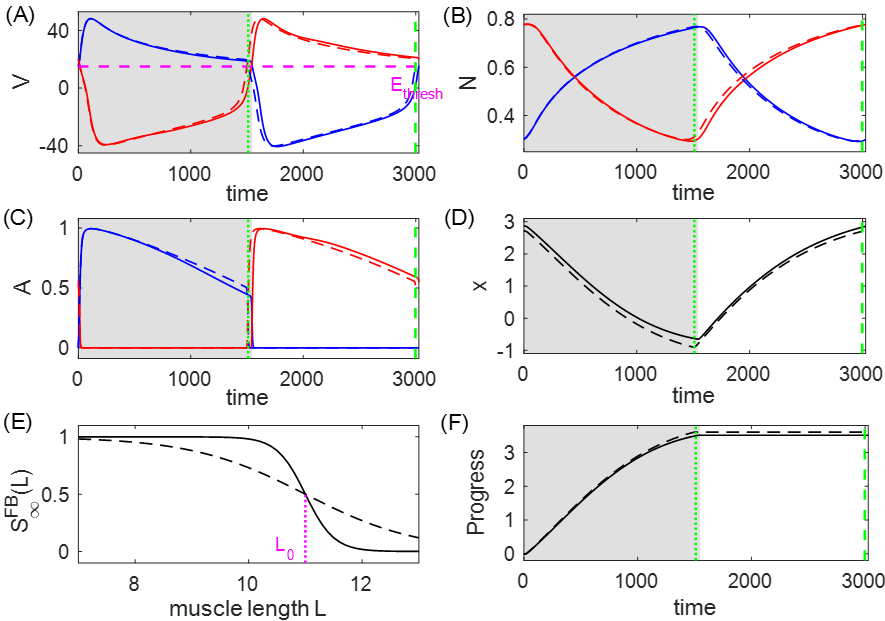}
\caption{\label{fig:Yu_slope_per} Comparison of the dynamics of two systems controlled by the ID mechanism with $L_0=11$ and $L_\text{slope}=0.5$ (solid traces) versus  $L_\text{slope}=2$ (dashed traces).
The gray shaded region denotes the powerstroke phase of the first (steeper slope) system ($T_0^\text{ps}=1547$); the green vertical dotted line represents the end of the powerstroke phase for the second system ($T_0^\text{ps}=1511$), and the green vertical dashed line represents the end of its full cycle.
During the powerstroke phase, the less steep synaptic activation $S^\text{FB}_\infty$ (panel E) reduces the feedback inhibition to $V_1$, leading to greater activation of the neuron driving the powerstroke (``neuron 1"), and earlier transition into the recovery phase (panel A).
The activation of muscle 1 becomes stronger due to the larger $V_1$ (panel C), which results in a stronger force being generated by muscle 1.  
This enhanced force pulls the object forward more quickly, especially around the end of power stroke (panel D).
Due to the faster movement, the object makes greater progress in an even shorter time (panel F), thereby outperforming the system with the more steep $S_\infty^\text{FB}$ curve.
} 
\end{figure}

\paragraph{The performance patterns in the decreasing and increasing feedback mechanisms are qualitatively reversed with respect to the steepness of $S_\infty^\text{FB}$.} 
For example, as $S_\infty^\text{FB}$ changes from constant to approaching a Heaviside step function, the performance of the II mechanism with $L_0=11$ (red dots in Fig.~\ref{fig:Yu_per_sen}A bottom) monotonically increases while the performance of the corresponding ID mechanism (red + in Fig.~\ref{fig:Yu_per_sen}A bottom) decreases monotonically.
Figure~\ref{fig:Yu_slope_per} illustrates two systems controlled by the ID mechanism with $L_0=11$ but different $L_\text{slope}$ values.
When $S_\infty^\text{FB}$ is more shallow (dashed), the inhibitory feedback current to cell 1 is less intense, due to the smaller synaptic activation over the contraction regime of muscle 2 (panel E).
This makes neuron 1 (the neuron driving the powerstroke) more active, switching earlier to the inhibited state, and thus giving a shorter duration power stroke (panel A).
The progress of the limb over the shorter powerstroke phase is however larger, which is opposite to the excitatory situation in Fig.~\ref{fig:Yu_ex_in_per}.
This outcome occurs because the effect of the duration decrease is not comparable to the effect of the faster velocity around the end of power stroke.
The progress velocity over the powerstroke phase is controlled by the force generated by muscle 1, which becomes stronger due to the increased muscle activation $A_1$ (panel C). 
Therefore, by decreasing the steepness of $S_\infty^\text{FB}$ for the ID mechanism, the muscles act on the limb in a stronger and faster fashion, leading to the enhanced performance.
In contrast, by applying a similar analysis, we observe that the corresponding II mechanism, for which the muscle-stretch activated feedback current gives more inhibition to the active neuron 1, has the opposite effect (not shown).
This example illustrates how two mechanisms with the opposite activating property of the sensory feedback can have qualitatively distinct performance changes when the sensitivity of the feedback pathway to the biomechanics is varied.

\begin{figure}[htb]
\centering
\includegraphics[width=8cm]{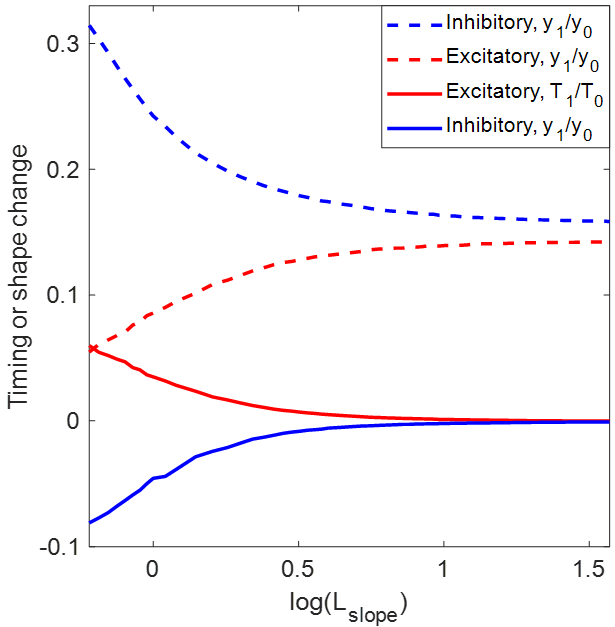}
\caption{\label{fig:Yu_sen_components} Effects of the increased load on the period ($T_1/T_0$, solid) and on the progress ($y_1/y_0$, dashed) of the trajectories for the ED systems (red) and ID systems (blue) with $L_0$ fixed at $L_0=9$ and $L_\text{slope}$ varied over $(0.6, 40)$. 
(Note this is a subset of the ranges plotted in Fig.~\ref{fig:Yu_per_sen}, limited to the region in which sensitivity varies monotonically.)
The absolute difference between the two effects accounts for the sensitivity of the system.
In the excitatory system, the positive $T_1/T_0$ and $y_1/y_0$ indicate that the perturbed system spends longer time to make larger progress.
The decrease of $L_\text{slope}$ reduces $y_1/y_0$ while $T_1/T_0$ is increased, so the sensitivity decreases, until perfect robustness is attained at $L_\text{slope}\approx0.6$, i.e.~where the two red curves meet each other.
In the inhibitory system, $T_1/T_0$ is negative, indicating that the perturbation shortens the working period.
The effects on the progress and period both become stronger with the decrease of $L_\text{slope}$, so the system becomes more sensitive to the perturbation.} 
\end{figure}

\paragraph{The sensitivity profiles of the excitatory and inhibitory systems to an applied load are qualitatively reversed with respect to the steepness of the synaptic activation function.} 
Consider the two contralateral-decreasing mechanisms with $L_0=9$ (see the two blue dot curves in Fig.~\ref{fig:Yu_per_sen}A).
As $L_\text{slope}$ becomes smaller, the sensitivity of the ED mechanism first decreases to almost 0 (perfect robustness) and then increases, whereas the sensitivity of the corresponding ID mechanism first increases to the maximum and then decreases.
Recall that following equation~\eqref{eq:sensitivity_formula2}, the sensitivity  can be written as
$$S=\left|\frac{\partial Q}{\partial\kappa}(\kappa_0)\right|=Q_0\left|\frac{y_1}{y_0}-\frac{T_1}{T_0}\right|.$$
This expression contains two terms --- $y_1/y_0$ accounts for the effect of the perturbation on the system progress (shape), while  $T_1/T_0$ accounts for the effect on the period (timing).
To the extent that the two effects are large and of opposite signs, the system becomes more sensitive; in contrast, the cancellation of the two effects leads to a robust system.
Figure~\ref{fig:Yu_sen_components} shows the two effects in the ED mechanism (red) and ID mechanism (blue) with $L_0=9$ and $L_\text{slope}$ varied over $(0.6,40)$.
We observe that the stronger mechanical load exerts positive effects on both the timing and shape for the system with the excitatory sensory feedback mechanism, by prolonging the limit cycle period and increasing the progress of the limb (positive red curves), but the shape effect is more profound.
As $L_\text{slope}$ decreases, the improvement in the progress becomes less significant, while the timing changes more dramatically, which leads to a reduction in the sensitivity.
The sensitivity minimum is attained when the two effects completely offset each other at $L_\text{slope}\approx0.6$, where the system is most robust against the perturbation.
In contrast, the perturbation allows the system with the inhibitory feedback to make larger progress in a shorter time (negative blue solid and positive blue dashed); this possibility was mentioned in the discussion of Fig.~\ref{fig:Yu_slope_per}.
Although making greater progress in a shorter time in response to perturbation improves performance, it is not beneficial to the robustness of the system in terms of maintaining performance homeostasis.
Moreover, as $L_\text{slope}$ decreases, both effects become stronger, inducing the system to be even more sensitive.
Apart from this example, the excitation-inhibition property of the sensory feedback in this model always offers qualitatively reversed sensitivity patterns when we consider the steepness variation of the feedback activation curve.
The variational analysis serves as a tool to examine the coordinated effects of perturbation on the trajectory geometry and timing, and to identify mechanisms with superior robustness. 

\begin{figure}[htb]
\centering
\includegraphics[width=9cm]{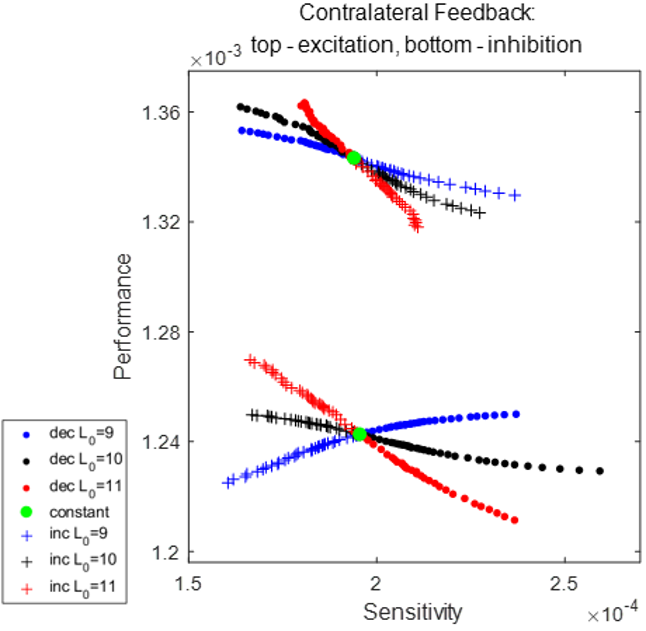}
\caption{\label{fig:Yu_per_sen_zoomin} Detail from Fig.~\ref{fig:Yu_per_sen}A, expanded to show the region of approximate linearity.
Colors as in Fig.~\ref{fig:Yu_per_sen}A.
The steepness parameter $L_\text{slope}$ of the synaptic feedback activation function $S_\infty^\text{FB}$ is varied over $(2,\infty)$.
Correspondingly, the gain of the sigmoid ranges from zero to 0.5.
With the small gain, the sigmoid $S_\infty^\text{FB}$ is almost linear over the working range of the muscle length, and the performance-sensitivity curve for each case changes approximately linearly with respect to $L_\text{slope}$.} 
\end{figure}

\paragraph{When $S_\infty^\text{FB}$ is approximately linear over the working range of the muscles and limb, the performance-sensitivity curve changes approximately linearly with $L_\text{slope}$.}
As an illustration, Figure~\ref{fig:Yu_per_sen_zoomin} zooms in on the patterns shown in Fig.~\ref{fig:Yu_per_sen}A with  $L_\text{slope}$ varied among $(2,\infty)$, over which $S_\infty^\text{FB}$ is approximately linear within the possible range of muscle length.
On this scale, a fundamental ambiguity arises on account of the dual goals of performance and robustness.  
Without specifying a relative weighting between these two quantities, there is no well justified way to choose which of the three traces in the upper ensemble, all moving up and to the left, are preferred.  
The red, black, and blue traces all simultaneously increase performance while decreasing sensitivity, relative to the constant feedback case (green dot).
Among the lower ensemble, the blue curve can be rejected as its components exhibit a tradeoff: the robustness is enhanced with decreased performance.
Moreover, within the linear regime, the performance and sensitivity can both be improved indefinitely by increasing the feedback gain.  
The possibility of disambiguating the two effects and finding a globally optimal solution arises only when the parameters are varied enough for nonlinear effects to come into play, as shown next.

\paragraph{The nonlinearity of the synaptic feedback activation function allows the existence of optimal combinations of performance and sensitivity.}
As the linear regime of $S_\infty^\text{FB}$ extends beyond the working range of muscle length and the curvature of the sigmoid $S_\infty^\text{FB}$ becomes more pronounced, some sharp turning points arise in the performance-sensitivity curve of each feedback mechanism, either in terms of performance or sensitivity or both.
In general, along a continuous curve in the (sensitivity $S$, performance $Q$) plane, indexed by a parameter $\mu$ (here, the sigmoid steepness parameter $L_\text{slope}$), there will be an optimal region marked at one end by the condition $(\partial S/\partial\mu=0, \partial Q/\partial\mu\not=0)$ and at the other end by $(\partial S/\partial\mu\not=0, \partial Q/\partial\mu=0)$.  
Depending on the relative weight given to $S$ and $Q$, the optimal value of $\mu$ will place the system somewhere within this segment of the $(S(\mu), Q(\mu))$ curve.
In the case of the HCO system, the optimal curve among those tested ($L_0=9$, top blue dots in Fig.~\ref{fig:Yu_per_sen}A) makes a hairpin turn in the upper left corner of the plot (blue arrow), leading to relatively unambiguous identification of the optimal value of the slope parameter $L_\text{slope}$.
In such cases, these optimal points, or narrowly identified optimal regions, indicate the possibility of well-defined simultaneous optima in both of the performance and sensitivity patterns.
Thus our results suggest the possibility to realize these joint goals through adjusting the structure of the sensory feedback mechanisms more broadly.

The preceding observations have provided several insights that could support the design of sensory feedback pathways, such as the selection of excitatory versus inhibitory feedback currents, and activation versus inactivation with muscle contraction, as well as the shape of the feedback activation curve, to promote efficiency and robustness in other more realistic HCO-motor models with analogous configurations.

\section{Application: Markin hindlimb model with imposed slope}
\label{sec:Markin model application}

\begin{figure}
\centering
\includegraphics[width=7.5cm]{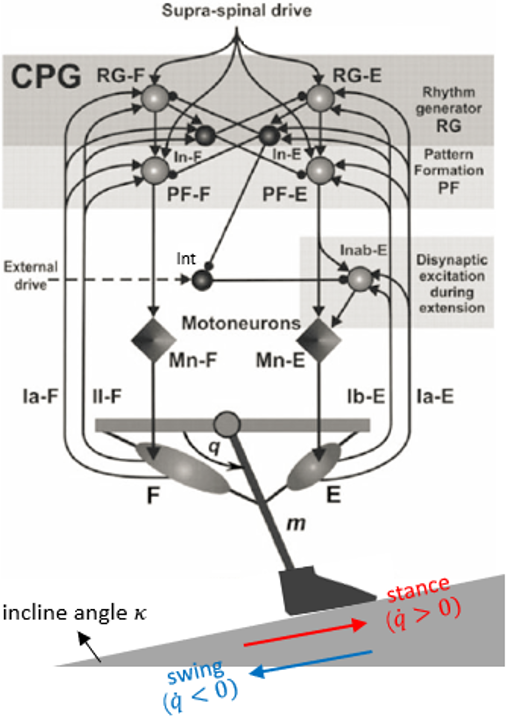}
\caption{\label{fig:Markin_sketch} 
Schematic of Markin hindlimb model with imposed slope, adapted from \cite{markin2010afferent}.
The spinal CPG, consisting of Rhythm Generator (RG) and Pattern Formation (PF) half-center oscillators, as well as Interneurons (In-E/F), receives tonic supra-spinal drive, and generates alternating activation driving corresponding flexor and extensor Motor neurons (Mn).
An additional circuit including more interneurons (Int and Inab-E) is incorporated to provide disynaptic excitation to Mn-E.
The Mn cells activate antagonistic flexor and exentors muscles, which control the motion of a single-joint limb.
Sensory feedback from muscle activation, which comes in three types (Ia, Ib, and II), provides excitation to the ipsilateral neurons of the CPG.
Inhibitory connections end with a round ball, and excitatory connections end with a black arrow. 
The limb stands on the ground with slope specified by $\kappa$, which we take as our perturbation parameter.
During a single locomotion cycle, the ground reaction force is active on the limb in the stance phase where the limb velocity $\dot{q}$ is positive (red arrow) but not in the swing phase where $\dot{q}<0$ (blue arrow).
} 
\end{figure}

The second closed-loop powerstroke-recovery example we consider is based on a neuromechanical model proposed by \cite{markin2010afferent}, as sketched in Fig.~\ref{fig:Markin_sketch}.
The model consists of a spinal central pattern generator controlling the movement of a single-joint limb.
The CPG sends output via efferent activation of two antagonist (flexor and extensor) muscles to a mechanical limb segment, which in turn generates afferent feedback signals to the CPG.
The model system performs rhythmic locomotion, comprising a \emph{stance} phase, during which the limb is in contact with the ground, and a \emph{swing} phase, during which the limb moves without ground contact.
Thus the system falls within the class of powerstroke-recovery systems.

Everyday experience teaches us that normal walking movements are robust against gradual changes in the slope of the terrain, although the detailed shape and timing of the limb trajectory changes as one ascends (or descends) a steeper or shallower incline.
As a proxy for this form of parametric perturbation of rhythmic walking movements, we extend Markin et al's model to include an \emph{incline} parameter, simulating the effects of a change in the ground slope.  
We use this parameter in our analysis of performance and sensitivity, to define the sensitivity of the system's progress relative to the external substrate (the ground) in response to this environmental change.

\subsection{The equations of the Markin model}

Following \cite{markin2010afferent}, we model the central pattern generator as a multi-layer circuit, consisting of a half-center rhythm generator (RG) containing flexor neurons (RG-F) and extensor neurons (RG-E).  
These RG neurons project to pattern formation (PF) neurons (PF-F and PF-E, respectively) and to inhibitory interneurons (In-F and In-E) which mediate reciprocal inhibition between the flexor and extensor half-centers.
Given sufficient tonic supra-spinal drive, the CPG generates rhythms of alternating activation of flexor and extensor neurons, and the output of PF neurons induces alternating activity in flexor and extensor motor neurons (Mn-F and Mn-E).
An additional circuit of interneurons (Int and Inab-E) provides a disynaptic pathway from the extensor side to Mn-E.

The dynamics of the RG, PF, and Mn neurons are each described by two first-order ordinary differential equations, governing each cell's membrane potential $V_i$ and the slow inactivation gate $h_i$ of a persistent sodium current:
\begin{equation}
\label{eq:Markin_equation_RG}
    \begin{split}
        C\frac{dV_i}{dt}=&-I_\text{NaP}(V_i,h_i)-I_\text{K}(V_i)-I_\text{Leak}(V_i)-I_\text{SynE}(V_i)-I_\text{SynI}(V_i),\\
        \frac{dh_i}{dt}=&(h_\infty(V_i)-h_i)/\tau_h(V_i).
    \end{split}
\end{equation}
Here, $I_\text{NaP}$, $I_\text{K}$, and $I_\text{Leak}$ refer to the persistent sodium current, potassium current, and leak current, respectively, described by
\begin{align*}
    I_\text{NaP}(V_i,h_i)&=g_\text{NaP}m_\text{NaP}(V_i)h_i(V_i-E_\text{Na}),\\
    I_\text{K}(V_i)&=g_\text{K}m^4_\text{K}(V_i)(V_i-E_\text{K}),\\
    I_\text{Leak}(V_i)&=g_\text{L}(V_i-E_\text{L}).
\end{align*}
Excitatory and inhibitory currents to neuron $i$ are respectively represented by $I_\text{SynE}(V_i)$ and $I_\text{SynI}(V_i)$, given by
\begin{equation}
\label{eq:Markin_currents}
\begin{split}
    I_\text{SynE}(V_i)&=g_\text{synE}(V_i-E_\text{synE})\left(\sum_{j}a_{ji}f(V_j)+c_id+\sum_{k}w_{ki}s_k\text{fb}_k\right),\\
    I_\text{SynI}(V_i)&=g_\text{synI}(V_i-E_\text{synI})\sum_{j}b_{ji}f(V_j).
\end{split}
\end{equation}
The nonlinear function $f$ describes the output activity of neuron $j$, defined to be
\begin{align}
\label{eq:neuron_output_f}
   f(V_j)=\left\{\begin{aligned}
&1\Big/ \left(1+\exp{\left(-\frac{V_j-V_{1/2}}{k}\right)}\right),\quad&V_j\geq V_\text{th},\\
&0,\quad&\text{otherwise},
\end{aligned}\right.
\end{align}
where $V_\text{th}$ is the synaptic threshold.
Parameter $a_{ji}$ defines the weight of the excitatory synaptic input from neuron $j$ to neuron $i$, while $b_{ji}$ defines the weight of the inhibitory input from $j$ to $i$; $c_i$ represents the weight of the excitatory drive $d$ to neuron $i$; $w_{ki}$ defines the synaptic weight of afferent feedback $\text{fb}_k$ to neuron $i$, with the feedback strength $s_k$.
Details on the feedback terms $\text{fb}_k$ will be provided at the end of this section. 

The dynamics of the interneurons, In-F, In-E, Int, and Inab-E, are each described by a single first-order equation:
\begin{equation}
\label{eq:Markin_equation_In}
     C\frac{dV_i}{dt}=-I_\text{Leak}(V_i)-I_\text{SynE}(V_i)-I_\text{SynI}(V_i),
\end{equation}
where the currents are in the same form as above. 
As shown in Fig.~\ref{fig:Markin_sketch}, the source of excitatory inputs $I_\text{SynE}$ to these interneurons comes from RG, supra-spinal drive, and  sensory feedback. 
Note that In-E and In-F in particular do not receive any inhibitory input  or excitatory supra-spinal drive, so the right-hand side of their voltage equation has $I_\text{SynI}=0$ and $c_i=0$. 

The motor neurons, Mn-F and Mn-E, respectively activate two antagonistic muscles, the flexor (F) and extensor (E), controlling a simple single-joint limb.
The limb motion is described by a
second-order differential equation:
\begin{equation}
\label{eq:Markin_phases}
    I\ddot{q}=\frac{1}{2}mgl_s\cos{q}-b\dot{q}+M_F(q,\dot{q},V_\text{Mn-F},t)+M_E(\pi-q,-\dot{q},V_\text{Mn-E},t)+M_\text{GR}(q,\kappa),
\end{equation}
where $q$ represents the angle of the limb with respect to the horizontal. 
The first term accounts for the moment of the gravitational force; $M_F$ and $M_E$ are the moments of the muscle forces; $M_\text{GR}$ denotes the moment of the ground reaction force which is active only during the stance phase, given by 
\begin{align}
\label{eq:M_GR}
M_\text{GR}(q,\kappa)=\left\{\begin{aligned}
&-M_\text{GRmax}\cos{(q-\kappa)},\quad&\dot{q}\geq0\text{ (stance / power stroke)},\\
&0,\quad&\dot{q}<0\text{ (swing / recovery)}.
\end{aligned}\right.
\end{align}
Note that the stance (powerstroke) phase is defined when the limb angular velocity $v=\dot{q}$ is nonnegative, while the swing (recovery) phase occurs when the velocity is negative.
We expand on the original model from \cite{markin2010afferent} by introducing parameter $\kappa$ to describe the slope of the ground whereon the limb stands.  Thus $\kappa$ will play the role of the load parameter subjected to perturbations for this model. 

The feedback signals from the extensor and flexor muscle afferents provide excitatory inputs to the RG, PF, In, and Inab-E neurons.
Muscle afferents provide both length-dependent feedback (type Ia from both muscles and type II from the flexor) and force-dependent feedback (type Ib from the extensor).
Linear combinations of feedback terms  $\text{fb}_k\in\{\text{Ia-F, II-F, Ia-E, Ib-E}\}$, written as $\sum_{k}w_{ki}s_k\text{fb}_k$ in \eqref{eq:Markin_currents}, are fed into each side of the model --- Ia-F and II-F feedback go to the flexor neurons and Ia-E and Ib-E to
the extensor neurons.
The feedback terms are in the form
\begin{equation}
\label{eq:Markin_fb_eq}
\begin{split}    \text{Ia}&=k_\text{v\text{\uppercase\expandafter{\romannumeral1}a}}\text{sign}(v^m)|v_\text{norm}|^{\rho_v}+k_\text{dIa}d_\text{norm}+k_\text{nIa}f(V_\text{Mn})+C_\text{\text{Ia}},\\
    \text{II}&=k_\text{dII}d_\text{norm}+k_\text{nII}f(V_\text{Mn})+C_\text{II},\\
    \text{Ib}&=k_\text{FIb}F_\text{norm}.
\end{split}
\end{equation}
For more details about the mathematical formulations, functional forms, and parameter values of the entire model, see Appendix~\ref{app:Markin_model_details}.

\subsection{Analysis of the Markin model}

\begin{figure}
\centering
\includegraphics[width=14cm]{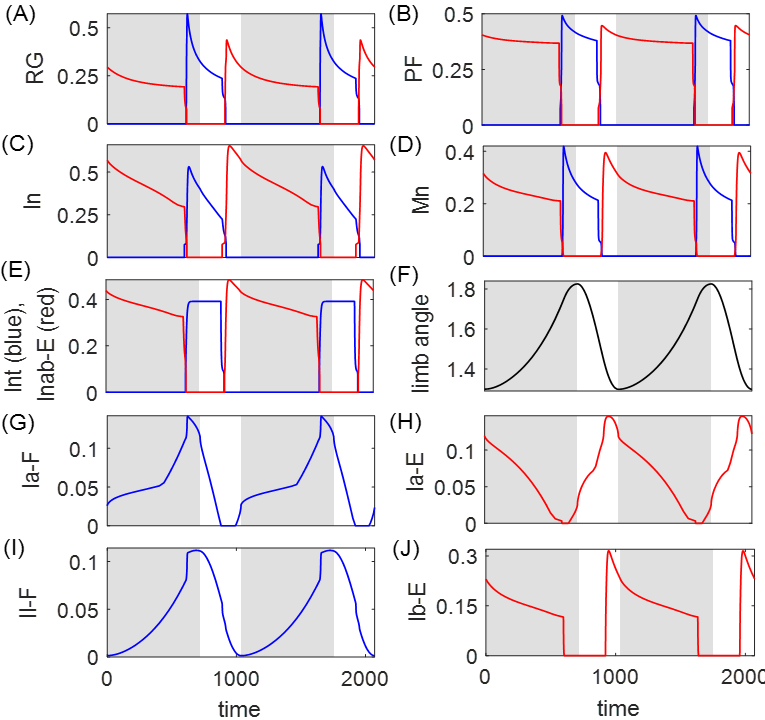}
\caption{\label{fig:Markin_sol_example} A typical solution for the Markin model, in (A) RG output, (B) PF output, (C) In output, (D) Mn output, (E) Int (blue) and Inab-E (red) output, (F) limb angle, (G) Ia-F feebback activity, (H) Ia-E feedback activity, (I) II-F feedback activity, and (J) Ib-E feedback activity.
The flexor side is colored by blue and the extensor side is colored by red.
The gray shaded region indicates the stance phase (the powerstroke) during which the limb angle is positive in velocity (panel F), and the white region indicates the swing phase.
The stance and swing phases are slightly shifted relative to the extensor and flexor active phases in the CPG.
Note that for rat hindlimb walking, the extensor muscle dominates the stance / powerstroke phase \cite{markin2010afferent}.
} 
\end{figure}

Figure~\ref{fig:Markin_sol_example} shows the time courses of the output of neurons $f(V_i)$, limb angle $q$, and feedback activity $\text{fb}_k$ in the default flat-ground system where $\kappa=0$.
Unlike the HCO model, where the active states of neurons overlap with the powerstroke and recovery phases, here the extensor and flexor active phases are slightly shifted relative to the stance and swing phases \cite{spardy2011dynamical}.
The excitatory feedback of types Ia and II increase during the silent phase of the associated neuron receiving the signal, reaching a peak just before the target neuron becomes active; these feedback signals then decreases during the active phase of the neuron.
In contrast, the Ib-E feedback signal, which is solely dependent on the extensor muscle force, is active only when the extensor neurons are active. It remains low until the onset of activation in the extensor units, at which point it jumps up to a high level and then recedes.  
The different types of the feedback pathways induce distinct performance-sensitivity patterns, as we will discuss in \S\ref{subsec:per_sen_markin}.

\begin{figure}[htb]
\centering
\includegraphics[width=14cm]{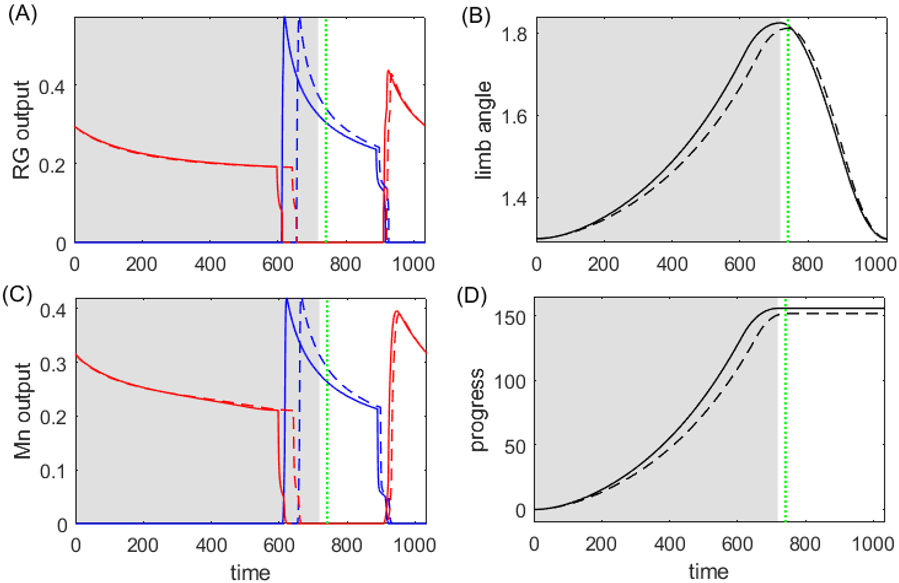}
\caption{\label{fig:Markin_compare_pert_unpert}
Comparison of time series of (A) output of RG neurons, (B) limb angle, (C) output of Mn neurons, and (D) progress, for the unperturbed trajectory (solid, same as Fig.~\ref{fig:Markin_sol_example}) and perturbed trajectory (dashed) with ground slope $\kappa_\epsilon=0.01$.
Both are initiated at the start of their respective stance phase.
The perturbed trajectory is uniformly time-rescaled by $t_\epsilon=\frac{T_0}{T_\epsilon}t$, where $T_0=1035$ and $T_\epsilon=1104$.
The shaded regions represent the stance phase for the unperturbed case, whereas the vertical green dotted lines represent the onset of the swing phase for the perturbed case, which is delayed due to the steeper ground. 
Compare Fig.~\ref{fig:Yu_iSRC_example}.} 
\end{figure}

For this model, we define the performance to be the average distance the limb moves along the ground during the stance phase, given by
\begin{align}
\label{eq:performance_Markin}
    Q(\kappa)=\frac{l_s}{T_\kappa}\int_0^{T_\kappa^\text{st}}-\frac{d}{dt}\left(\cos{(q(\gamma_\kappa(t))-\kappa})\right)\,dt.
\end{align}
Here $T_\kappa^\text{st}$ denotes the stance duration and $l_s$ is the limb length.
Figure~\ref{fig:Markin_compare_pert_unpert} compares the time series of the perturbed system subjected to a small change in the ground slope ($\kappa_\epsilon=\epsilon=0.01$) with the unperturbed system.
As indicated by the green dotted vertical line, the perturbation prolongs the stance phase and delays the transition to the swing phase, in that the system decelerates because of the steeper ground slope (panel B).
Consequently, the progress becomes smaller (panel D), and the performance is further deteriorated due to a longer period ($Q_0=0.1506$, $Q(\epsilon)=0.1375$).
Note that this model and the HCO model have different conditions for transitioning between the powerstroke and recovery phases.
In the HCO model the transition is determined by the neuron activation, while in the biophysically-grounded Markin model the transition is determined by the mechanical condition $\dot{q}=0$.
This difference accounts for the opposite responses of the transition timing to analogous environmental challenges: when increasing the difficulty of the task the HCO powerstroke-to-recovery transition moves earlier, while the Markin model's transition is delayed.
Put another way, the powerstroke of the perturbed HCO system contracts relative to the net period, while in the Markin model it expands (cf.~Fig.~\ref{fig:Yu_iSRC_example}).

\begin{figure}[htb]
\centering
\includegraphics[width=11cm]{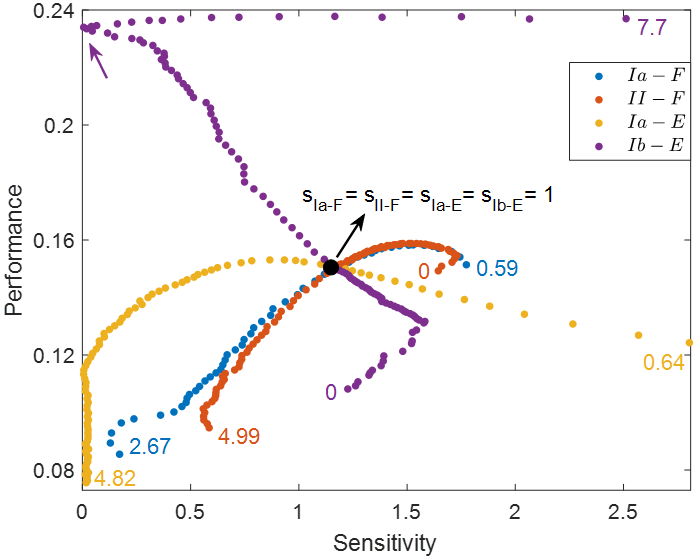}
\caption{\label{fig:Markin_per_sen} Performance-sensitivity patterns of the Markin model for different feedback strengths.
The black dot represents the default system with the normal feedback strength (all $s_k=1$).
In each curve, the strength for only one feedback pathway is varied, with the other three strengths fixed at one.
The number at each end indicates the maximal or minimal strength allowing for stable progressive locomotor oscillations.
The purple arrow marks the most advantageous configuration among those tested.
} 
\end{figure}

In the following, we study how the performance-sensitivity pattern is affected by the strength of each afferent feedback pathway, represented by $s_k$ where $k\in\{\text{Ia-F, II-F, Ia-E, Ib-E}\}$ in \eqref{eq:Markin_currents}.
Figure~\ref{fig:Markin_per_sen} shows the patterns as one of the four feedback strengths is varied, with the other three strengths fixed at the normal strength $s_k=1$.  
Before we discuss the patterns in \S\ref{subsec:per_sen_markin}, note that the possible range of each strength allowing for stable progressive locomotor oscillations is remarkably different:
$$s_\text{Ia-F}\in[0.59,\,2.67],\quad s_\text{Ia-E}\in[0.64,\,4.82],\quad s_\text{II-F}\in[0,\,4.99],\quad s_\text{Ib-E}\in[0,\,7.68].$$
In particular, the system can maintain stable oscillations upon cutting off the II-F or Ib-E feedback pathway, but cannot sustain oscillations without the Ia-type feedback.
We finish this subsection by examining the mechanism underlying the failure of movement as the strength parameter is out of the range.

\begin{figure}[htb]
\centering
\includegraphics[width=14cm]{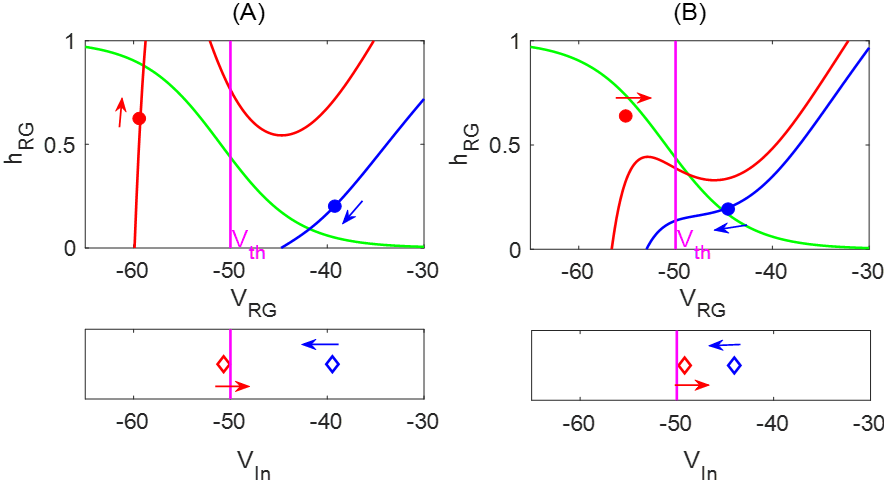}
\caption{\label{fig:Markin_nullcline} Escape of In triggers CPG transitions. 
\textbf{Top}: Nullclines and positions of RG neurons (dots) in the $(V_\text{RG}, h_\text{RG})$ phase plane.
\textbf{Bottom}: Voltages of In neurons (diamonds) at corresponding times.
Blue indicates flexor $V$-nullcline, voltage and/or $h$ values.
Red indicates extensor nullcline and values.
The $h$-nullcline is shown in green. 
The vertical magenta line represents the synaptic voltage threshold $V_\text{th}$.
Arrows mark the direction the neurons are moving.
\textbf{(A)}: When the extensor neurons are at the inhibited state, the $h$-nullcline intersects the inhibited $V$-nullcline of RG-E at the left branch. 
Excitation from feedback allows In-E to reach the threshold first, independently of RG-E.
\textbf{(B)}: When In-E jumps above the threshold it begins to inhibit RG-F, which raises the $V$-nullcline of RG-F.
The downstream In-F, receiving less excitation from RG-F, thus reduces its voltage and gives less excitation to RG-E, which lowers the $V$-nullcline of RG-E.
As a result, RG-E lies above the left knee of its $V$-nullcline, so it jumps across the threshold and becomes active, switching the dominance in the CPG (not shown).
} 
\end{figure}

The Markin model system features fast-slow dynamics \citep{rubin2002geometric}, as the persistent sodium inactivation time constant $\tau_h(V_i)$ in \eqref{eq:Markin_equation_RG} is large over the relevant voltage range, so $h$ evolves on a slower timescale than $V_i$.
The activity of neuron $i$ can be therefore determined from the location of the intersection of its nullclines in the $(V_i,h_i)$ phase plane.
To illustrate the mechanism underlying the transition between phases, Fig.~\ref{fig:Markin_nullcline} shows the nullcline configuration and the corresponding positions of RG and In neurons in the default system around the CPG transition.
Starting from the extensor-inhibited state (panels A), because the $V$-nullcline and $h$-nullcline of RG-E intersect at a silent stable fixed point, RG-E cannot escape from the silent state and trade dominance on its own.
However, In-E, due to the increasing excitatory inputs from Ia-E and Ib-E feedback (Fig.~\ref{fig:Markin_sol_example}), is able to cross the synaptic threshold first and begin to inhibit RG-F (panels B). 
This inhibition raises the $V$-nullcline of RG-F, followed by the decrease of RG-F voltage, weakening its excitation to the downstream In-F.
This reduction of excitation results in the decrease of In-F voltage.
Consequently, RG-E receives less inhibition from In-F, lowering the $V$-nullclines of RG-E, such that the critical point moves to the middle branch of its $V$-nullcline, and hence, RG-E can reach the left knee and jump to the right branch.
This transfer of active and silent states indicates that the In cells and excitatory sensory feedback dominate the transitions in the CPG.
In the case without sufficiently strong feedback inputs to the silent In cell, the ipsilateral RG neuron will become deadlocked in silence, and thus the locomotion will fail.

\begin{figure}
\centering
\includegraphics[width=13cm]{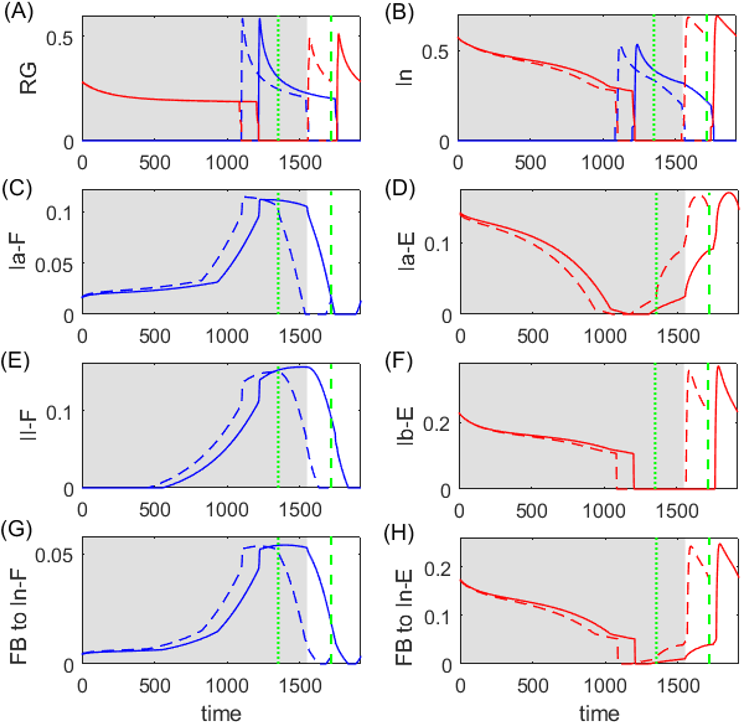}
\caption{\label{fig:Markin_limit} 
Effects of reduced sensory feedback gain $s_\text{Ia-F}$ to In cells near the collapse point.
Dashed curves: $s_\text{Ia-F}=0.63$.
Solid curves: $s_\text{Ia-F}=0.6$.
For all plots, red traces indicate extensor and blue traces indicate flexor quantities.
The vertical green dotted and dashed lines indicate the end of stance and swing phases for the case $s_\text{Ia-F}=0.63$.
The shaded and white regions represent the stance and swing phases, respectively, for the case $s_\text{Ia-F}=0.6$, both of which are significantly prolonged.
Further reduction of $s_\text{Ia-F}$ below 0.59 leads to stalling (convergence to a stable fixed point).
Panels G and H plot the weighted feedback inputs ($\sum_{k}w_{ki}s_{k}\text{fb}_k$) fed into each In cell.
Although the feedback magnitude does not differ much between the two cases, the time to reach the peak is significantly different, which affects the extensor-flexor transition and the following stance-swing transition.
} 
\end{figure}

Disentangling the effects of changing the strength of an afferent feedback pathway is nontrivial, because of the multiplicity of pathways impacting the activities of many neurons and limb within the system.
For instance, Fig.~\ref{fig:Markin_limit} shows the feedback and neuron dynamics in the cases $s_\text{Ia-F}=0.63$ (dashed) and $s_\text{Ia-F}=0.6$ (solid), which is approaching the minimal allowable strength of Ia-F.
Varying the strength of the Ia-F pathway alone leads to a chain of changes in all system components including the extensor feedback, and significantly impacts the transition timing.
Although the magnitude of the feedback signals does not differ much in the two systems, yet the system with smaller $s_\text{Ia-F}$ accumulates feedback excitation more slowly.
This circumstance is due to the interplay of spatial and timing measures of distance in the fast-slow dynamical systems \citep{terman1998dynamics,rubin2002geometric,yu2023sensitivity}, in which a small spatial difference translates into a significant temporal extension.
Decreasing the feedback strength $s_\text{Ia-F}$ further (below 0.59) makes it impossible for the extensor feedback to jump up; 
consequently the In-E neuron will not be facilitated to escape from the inhibited state (not shown) and the movement will stall.
This outcome may seem counterintuitive, since reducing the flexor feedback strength should make the flexor In more difficult to escape; however, its direct effect on the mechanics in turn influences the extensor feedback to an analogous (Ia-E) or even more significant (Ib-E) extent (see~equations~\eqref{eq:Markin_fb_eq}).
We observe similar situations beyond the nonzero extrema of other feedback strengths, in which either In-F or In-E fails to escape from the silent state and compromises the whole locomotor oscillation.

\subsection{Performance and sensitivity of the Markin model}
\label{subsec:per_sen_markin}

When analyzing the abstract HCO model, we considered both ipsilateral and contralateral feedback, both excitatory and inhibitory.
In contrast, each of the three sensory feedback pathways in the Markin model is ipsilateral and excitatory.  
Although the two models are quite different in their level of details, feedback variations, and the type of perturbation, we compare and contrast the two models as far as we can in the Discussion.
In this section, by varying the strength of each feedback pathway, we study the performance-sensitivity patterns of the Markin model as shown in Fig.~\ref{fig:Markin_per_sen} and yield the following observations.

\paragraph{Trade-offs between performance and sensitivity often occur as the afferent feedback strength changes.}

\begin{figure}[htb]
\centering
\includegraphics[width=14cm]{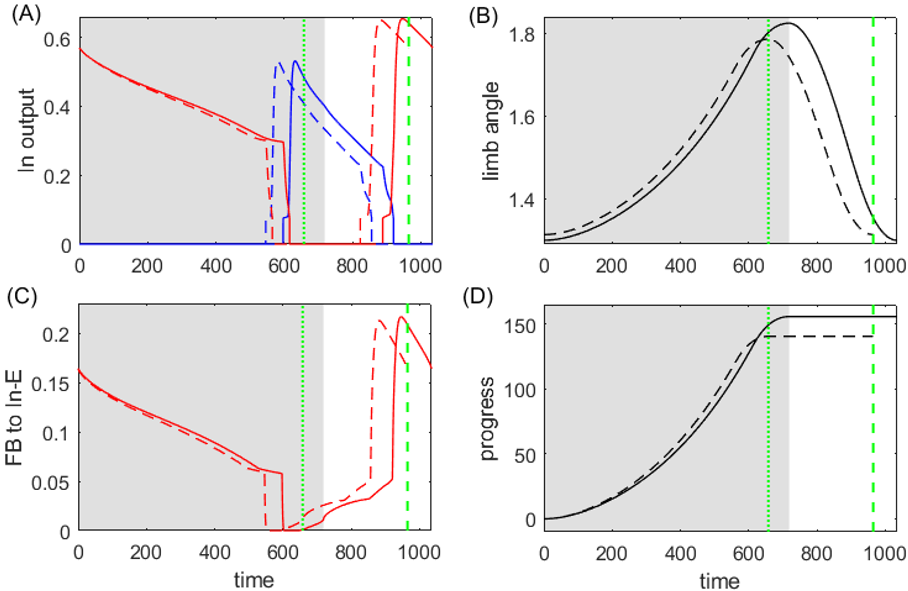}
\caption{\label{fig:Markin_compare_Iaf} Comparison of the dynamics of two systems with $s_\text{Iaf}=1$ (solid traces) versus $s_\text{Iaf}=1.1$ (dashed traces).
The other three feedback strengths are fixed at one.
The shaded region and white region denote the stance phase and swing phase, respectively, for the first system.
The vertical green dotted and dased lines represent the end of the stance phase and swing phase, respectively, for the second system, both of which are advanced due to the faster decay and accumulation of feedback excitation to In neurons (panel C).
The limb therefore makes less progress during the shorter stance phase (panel D).
Since the temporal decrease in the total period is less dominant than the spatial decrease in the progress, the performance (progress/period) of the system with the larger Ia-F strength is inferior to the default system (0.1457 vs 0.1505).
} 
\end{figure}

As the strength parameter is varied, the system shows a performance-sensitivity tradeoff when the curve either moves up and to right (performance improves while robustness decreases) or moves down and to left (robustness increases while performance declines).
Specifically, the system is not able to generate simultaneously efficient and robust movement with the whole flexor-feedback (Ia-F and II-F) variations as well as with the large Ia-E and small Ib-E variations.
For example, Fig.~\ref{fig:Markin_compare_Iaf} compares the performance of the default system to the system with a stronger Ia-F pathway ($s_\text{Iaf}=1.1$).
Similarly to Fig.~\ref{fig:Markin_limit}, the increase in the Ia-F strength advances the stance-swing transition, leading to a shortened working period and reduced progress.
Since the change in the shape is more significant than the change in the timing, the performance ($Q_0=y_0/T_0$) declines.
Note that in the HCO systems featuring either excitatory feedback or inhibitory feedback, the dominance between the shape and timing is reversed (cf.~Fig.~\ref{fig:Yu_ex_in_per}). 
These examples indicate the interplay of temporal and spatial effects in the performance measure.
Likewise, the sensitivity measure, given by \eqref{eq:sensitivity_formula1} or \eqref{eq:sensitivity_formula2}, also consists of two factors --- the timing response to the perturbation and the shape response.
Although the system with strengthened Ia-F feedback is inferior to the default system in terms of performance, it benefits from  enhanced robustness.
In contrast, weakening the Ia-F strength contributes to an improved performance at the cost of responding more sensitively to external perturbations.
With such a performance-sensitivity tradeoff, the system cannot simultaneously achieve the dual goals of efficiency and robustness, especially when varying the flexor-feedback pathway(s) alone.

\begin{figure}
\centering
\includegraphics[width=8cm]{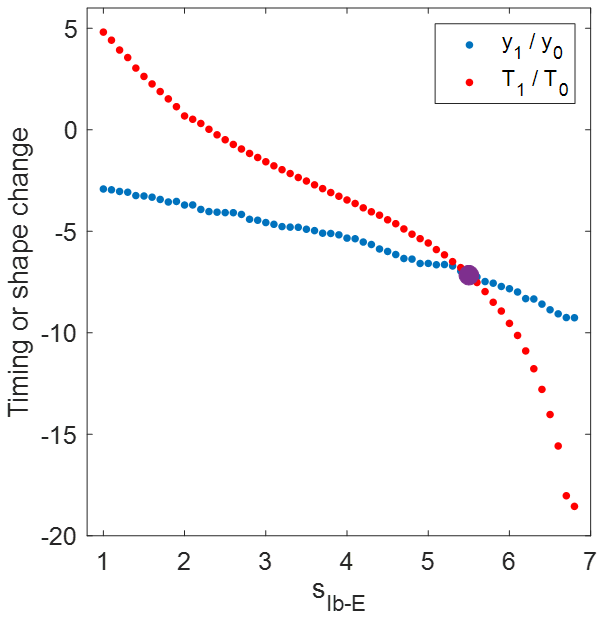}
\caption{\label{fig:Markin_sen_components} Effects of the Ib-E feedback strength $s_\text{Ib-E}\in(1,7)$ on the timing ($T_1/T_0$, red) and on the shape ($y_1/y_0$, blue) of the solution trajectories for the Markin model.
The purple dot represents the zero-sensitivity value, where $s_\text{Ib-E}=5.5$, which also corresponds to near-peak performance (cf.~purple arrow in Fig.~\ref{fig:Markin_per_sen}).} 
\end{figure}

\paragraph{Force-dependent feedback gain can optimize both performance and robustness simultaneously.}
We observe that the force-dependent feedback pathway (type Ib from the extensor) impacts performance and robustness very differently than the length-dependent feedback pathways (types Ia and II).
In contrast to the length-dependent feedback, the force-dependent feedback is only active during the active phase of the associated motor neuron, reflecting the fast-slow dynamics of the CPG (see equations~\eqref{eq:Markin_fb_eq} and Fig.~\ref{fig:Markin_sol_example}).
Notably, the performance-sensitivity tradeoff disappears with $s_\text{Ib-E}$ variations, and an optimal value of $s_\text{Ib-E}$ arises, giving the best combination of the pair $(S,Q)$ (purple arrow in Fig.~\ref{fig:Markin_per_sen}).
Around the optimal point, the performance saturates while the sensitivity changes dramatically, and at $s_\text{Ib-E}\approx5.5$, the sensitivity attains zero,  indicating that the system achieves infinitesimal homeostasis \citep{yu2022homeostasis} concurrent with peak  performance.
We verify this perfect robustness by plotting the coordinated effects of the slope perturbation on the system trajectory's shape and timing in Fig.~\ref{fig:Markin_sen_components} as $s_\text{Ib-E}$ varies over $(1,7)$.
For values of $s_\text{Ib-E}$ above the optimum, the robustness decreases dramatically.
Indeed, the underlying oscillation fails when $s_\text{Ib-E}$ reaches about 7.7, presumably due to a global bifurcation.
As the $s_\text{Ib-E}$ parameter approaches this point, the trajectory period and amplitude might be insensitive to the parameter variation but the inherent sensitivity to perturbations may dramatically increase. 
This circumstance is also observed in the HCO model when (subcritical) pitchfork bifurcations occur \citep{yu2021dynamical}.
On the contrary, with a strong Ia-E pathway (left corner of Fig.~\ref{fig:Markin_per_sen}), the system maintains robustness while its performance significantly decreases, which also appears around a (subcritical) Hopf bifurcation as shown in \cite{yu2021dynamical}.
Although a detailed bifurcation analysis of oscillation termination is beyond the scope of this paper, our observations demonstrate the difficulty of achieving the joint goals of high performance and low sensitivity in a realistic rhythmic physiological system. 
The variational method, by incorporating both temporal and spatial factors, offers the possibility of an analytic framework that may assist to identify the optimal sensory feedback control mechanism in regulating the system's efficiency and robustness.

\section{Discussion}
\label{sec:discussion}

\emph{Efficiency} and \emph{robustness} are important aspects of control systems in both engineering and biological contexts.
The application of conceptual and mathematical ideas from control theory in the physiological sciences has a long history \citep{baylis1966living,grodins1967mathematical,khoo2018physiological}. 
In motor control, viewed through the lens of systems engineering, the concepts of robustness and efficiency can help quantify different aspects of system performance underlying biological fitness.  
The roles of feedforward, reciprocal, and feedback motor pathways in contributing to robustness and efficiency must be investigated on a case-by-case basis. 
In many instances, control theorists have observed tradeoffs between efficiency and robustness, although the definitions of these terms may vary from one context to another 
\citep{kuo2002relative,boulet2007fundamental,ronsse2008control,vasconcelos2009stability,alfaro2010performance,sariyildiz2013performance}.
These authors describe a fundamental tradeoff between efficiency and robustness, in the sense that the system improves efficiency at the cost of becoming increasingly fragile to a slight change in any parameter.
Hence, an important aspect of controller design is to achieve an acceptable balance between the goals of increasing  performance versus maintaining robustness to external changes, when these objectives are in conflict.
Throughout this paper, we  interpret efficiency as quantifying how well a system performs, and robustness as quantifying the stability of efficiency with respect to parametric perturbations.

A remarkable feature of \emph{rhythmic} motor systems underlying a wide range of animals' physiological behaviors (crawling, breathing, scratching, swallowing) is the \emph{closed-loop} control structure, which integrates brain, body, and environment via sensory feedback \citep{ChielBeer1997TINS,diekman2017eupnea,lyttle2017robustness,chen2021bioinspired,korkmaz2021locomotion}.
In particular, feedback control plays a significant role in regulating the robustness and efficiency of these rhythmic movements.\footnote{The significance of sensory feedback may vary depending on the length scale of the animals, e.g.~in small insects versus large mammals \citep{SuttonSzczecinskiQuinnChiel2023PNAS}.}
Through afferent feedback pathways,  sensory information can communicate current demands, thereby allowing the system to update the corresponding movement on short time scales.
In this way, sensory feedback may improve performance; however, robustness remains a critical issue.
For example, \cite{kuo2002relative} and \cite{yu2021dynamical} discussed the fundamental properties of the trade-off between efficiency and robustness for a combined feedback–feedforward model for rhythm generation.  
These papers demonstrated that a system with pure feedback control, analogous to a chain reflex system, can compensate for unexpected disturbances, but shows poor robustness to imperfect sensors, and that the best trade-off requires a proper design incorporating both feedback and feedforward pathways.

In the present paper, we propose a unified framework to investigate how the architecture of sensory feedback influences the efficiency and robustness of rhythmic systems.
In the particular context of a general \emph{powerstroke-recovery} system, where the system repeatedly engages and disengages with the outside world, we measure the efficiency of the system, i.e.~the task performance, in terms of the physical progress of the system over the working period (or equivalently, the mean rate of progress).
This measure of the task performance is adopted from Lyttle et al.'s study of the feeding behavior of the sea slug \textit{Aplysia californica}, which considered mean seaweed consumption rate (cm/sec) \citep{lyttle2017robustness,wang2022variational}. This measure can be generalized to any powerstroke-recovery systems in which the progressive activity is important. 
We probe the robustness of the system by studying its ability to maintain task performance despite external perturbations. 
Following \cite{yu2022homeostasis}, we generalize the analytic formula for this robustness (or sensitivity) measure by using the infinitesimal shape response curve (iSRC) and the local timing response curve (lTRC) developed in \cite{wang2021shape}. These tools provide a mathematically grounded numerical quantification of the system's response to perturbations. 
By simultaneously studying the two objectives, we compare the performance-sensitivity patterns obtained in different feedback architectures, and identify optimal designs achieving the joint requirements of high performance and low sensitivity.

Here, we consider the control problem with two specific neuromechanical systems that produce powerstroke-recovery oscillations.
In the paradigmatic half-center oscillator (HCO)-motor model with an external load, adapted from \cite{yu2021dynamical}, 
we conduct a comprehensive analysis of different feedback mechanisms in the hopes that we uncover insights that may apply more broadly.
We study both contralateral and ipsilateral feedback architectures, both excitatory and inhibitory feedback modes, and both activating and inactivating pathways, in terms of their performance-sensitivity patterns.
First, we find that exchanging excitatory and inhibitory architectures reverses the effect of increasing feedback gain on the  sensitivity.
Second, we observe that the performance response is reversed depending on whether the feedback is activated or inactivated by muscle contractions. 
More interestingly, as the sigmoid activation of the feedback signal (as a function of limb position) becomes increasingly nonlinear over the working range of the limb, a well-defined simultaneous optimum in performance and sensitivity arises.
These findings may guide the selection of modeling frameworks to capture experimental observations or design aims for more realistic brain-body-environment systems with analogous architectures in future work.

In the hindlimb locomotor model \citep{markin2010afferent}, based on experimental measurements from spindle primary afferents in the cat, we modify the model by imposing a ground slope, and then manipulate the  gain parameters for each of four sensory feedback pathways.
We find that the force-dependent (type Ib-E) feedback can simultaneously optimize both performance and robustness, while the length-dependent (types Ia-E, Ia-F, and II-F)  feedback variations result in marked performance-versus-sensitivity trade-offs. 
Indeed, as discussed in \cite{yu2021dynamical}, the situation that the performance is increased by sacrificing the robustness is also observed in the abstract HCO-motor model when the system approaches a subcritical pitchfork bifurcation; and also, the situation that the robustness is improved at the cost of losing efficiency occurs when the system approaches a subcritical Hopf bifurcation. 

However, it is difficult to pursue further comparisons between the two models. 
In the HCO model, we consider only one feedback pathway, solely dependent on the muscle length.
In contrast, the locomotor model possesses multiple feedback channels, each of which relies on several components including muscle length, muscle velocity, motor neuron output, or muscle force.
As argued in \cite{katz2023conclusion}, although focusing on a specific model can advance understanding of the neural basis of behavior in species with similar developmental and physical constraints, doing so may overlook insights that could be obtained by considering convergent evolution as a framework leading to more general principles. 
Our HCO model can provide a mechanistic understanding leading to predictions about structure-function relationships, and allows us to compare different kinds of architectures that are ``equivalent" only at a more general level.
On the other hand the detailed Markin model is just one specific architecture where perhaps alternative architectures could have evolved to serve the same purpose. 
Thus, although the analogy between the HCO-motor and the Markin-motor model is limited, both models are worth studying from a conceptual point of view.
In particular, we note that the Ia- and II-type afferent feedback in  Markin's more realistic locomotor model is ipsilateral, excitatory, and increasing with muscle length; the performance-sensitivity pattern for this combination of elements nominally falls into the upper ensemble of the HCO model's ipsilateral patterns (Fig.~\ref{fig:Yu_per_sen}B, upper star traces).
In the HCO setting, these patterns are always advantageous over other patterns in terms of performance, and give rise to the only (performance, sensitivity) optimum.
This superiority could be interpreted as suggesting that our modeling analysis might be consistent with the action of natural evolution by which species adapt to environmental demands.

Apart from the restricted scope of feedback functions considered in our examples, it would be desirable to have greater insights into the way any form of sensory feedback shapes the attributes of motor control systems, both in terms of performance and in terms of robustness.
However, given the complexity and difficulty to ascertain sensory feedback signals experimentally, it is natural to restrict attention to monotonically increasing or decreasing sensory feedback signals, such as an excitatory or inhibitory conductance in a central pattern generator that increases or decreases steadily as a function of limb position, muscle stretch, muscle velocity, or other biomechanical variables.
A realistic example is the responses of cat spindle (group Ia and group II) afferents,  which monotonically change with muscle length and velocity signals, and tendon organ (group Ib) afferents, which scales in approximate proportion to muscle forces \citep{prochazka1998models,prochazka1999quantifying}.
Another example is the chemosensory pathway in the respiratory control system, which has been modeled by a sigmoidal relationship between the arterial partial pressure of oxygen and the conductance representing external drive to the CPG \citep{diekman2017eupnea}.
Moreover, biological signals are typically limited in range.
For instance, the firing rate of a neuron is bounded below by zero and typically bounded above by a maximal firing rate which is related, for example, to the neuron's absolute refractory period.  

The muscle dynamics of the HCO model was inspired in part by \textit{Aplysia}'s I2 muscle \citep{yu1999biomechanical}.
A nominal feeding model, featuring grasper-retraction (powerstroke) and grasper-protraction (recovery), has been well studied by Chiel and colleagues \citep{shaw2015significance,lyttle2017robustness,wang2022variational}. 
In their model, the proprioceptive feedback input is assumed to be affine linear with the grasper position ---  decreasing for the protraction neural pool while increasing for the retraction neural pool.
Depending on the grasper position and proprioceptive neutral position, the feedback signal can switch between excitation and inhibition during a single swallowing cycle.
In our HCO model, we only consider the same form of feedback functions for all neurons and the excitation-inhibition property does not change during the whole movement.
These differences point out an interesting and important direction for further study --- whether mixed inhibitory and excitatory feedback in a single pathway, or heterogeneous activating/inactivating feedback across different pathways ---  could provide  more realistic models for a broader range of specific motor systems.  

Finally, we apply a fast-slow decomposition analysis to the Markin model when discussing the failure of oscillations as the feedback gain parameter reaches the collapse point.
Although the stance and swing phases in the locomotion system do not differ much concerning their time scale,  it is nevertheless possible that in some powerstroke-recovery systems, the powerstroke, under a heavy load, could proceed on a slow timescale while the recovery could proceed on a fast timescale. 
In this case one might connect powerstroke-recovery systems to two-stroke relaxation oscillators as studied in depth by \cite{JelbartWechselberger2020Nonlin}, an interesting direction for future work.
Another avenue for future inquiry would be to consider the scaling of robustness and sensitivity structures, as studied here, in motor systems spanning a range of length and time scales \citep{SuttonSzczecinskiQuinnChiel2023PNAS}.
The approaches we use here should be applicable to these systems as well.

\section{Acknowledgments}

This work was supported in part by (1) National Institutes of Health BRAIN Initiative grant RF1 NS118606-01;
(2) the National Science Foundation under Grant No. DMS-1929284 while the authors were in residence at the Institute for Computational and Experimental Research in Mathematics in Providence, RI, during the program; and
(3) the Oberlin College Department of Mathematics.
The authors thank Hillel J.~Chiel and Yangyang Wang for helpful discussions concerning the manuscript.

\appendix

\section{HCO model details}
\label{app:Yu_model_details}

For completeness, we list here the full equations of the HCO model with an externally applied load, as introduced in Section \ref{sec:HCO model application} and originally proposed in \cite{yu2021dynamical}.
For cell $i,j=1,2$ and $j\neq i$,
\begin{equation*}
\begin{split}
C\frac{dV_i}{dt}&=I_\text{ext}-g_\text{L}(V_i-E_\text{L})-g_\text{Ca}M_\infty(V_i)(V_i-E_\text{Ca})-g_\text{K}N_i(V_i-E_\text{K})\\
&\qquad\quad-g_\text{syn}^\text{CPG}S_\infty^\text{CPG}(V_j)\left(V_i-E_\text{syn}^\text{CPG}\right)-g_\text{syn}^\text{FB}S_\infty^\text{FB}(L_{i,j})(V_i-E_\text{syn}^\text{FB}),\\
\frac{dN_i}{dt}&=\lambda_N(V_i)(N_\infty(V_i)-N_i),
\end{split}
\end{equation*}
where
\begin{align*}
S_\infty^\text{CPG}(V_j)&=\frac{1}{2}\left(1+\tanh\left(\frac{V_j-E_\text{thresh}}{E_\text{slope}}\right)\right),\\
S_\infty^\text{FB}(L_{i,j})&=\frac{1}{2}\left(1\pm\tanh\left(\frac{L_{i,j}-L_0}{L_\text{slope}}\right)\right),\\
M_\infty(V_i)&=\frac{1}{2}\left(1+\tanh\left(\frac{V_i-E_1}{E_2}\right)\right),\\
N_\infty(V_i)&=\frac{1}{2}\left(1+\tanh\left(\frac{V_i-E_3}{E_4}\right)\right),\\
\lambda_N(V_i)&=\phi_N\left(\cosh\left(\frac{V_i-E_3}{2E_4}\right)\right).
\end{align*}
The force of muscle $i$ is given by
$$F_i(t)=F_0\cdot a_i(t)\cdot LT(L_i(t)).$$
Here,
\begin{align*}
LT(L_i)&=-\frac{3\sqrt{3}}{1250}(L_i-1)(L_i-5)(L_i-15),\quad 5\leq L_i\leq15,\\
L_1&=10+x,\quad L_2=10-x,\quad-5\leq x\leq5,
\end{align*}
and
$$a_i(t)=g[A_i(t)-a_0]_+,\quad\quad0<a_i\leq1,$$
where
\begin{align*}
\frac{dA_i}{dt}&=\tau^{-1}\{U(V_i)-[\beta+(1-\beta)U(V_i)]A_i\},\\
U(V_i)&=1.03-4.31\exp{(-0.198(V_i/2))},\quad V_i\geq16.
\end{align*}
Note that in the $U$-equation, we leave implicit a conversion factor from mV to  Hz.
The limb position is controlled by
\begin{align*}
\frac{dx}{dt}&=\frac{1}{b}(F_2-F_1+r\kappa F_\ell),
\end{align*}
where
\begin{align*}
\label{eq:indicator_r}
r=\left\{\begin{aligned}
&1,\quad&\text{power stroke},\\
&0,\quad&\text{recovery}.
\end{aligned}\right.
\end{align*}
Table~\ref{table:HCO_parameters} specifies the parameter values used for simulations.
The simulation codes are available from \texttt{https://github.com/zhuojunyu-appliedmath/Powerstroke-recovery}.
Instructions for reproducing each figure and table in the paper are provided (see the README file at the github site).

\begin{table}
\caption{Parameter values of the HCO model}
\label{table:HCO_parameters}
\centering
\begin{tabular}{llllll}
\hline
Parameter&Value&Unit&Parameter&Value&Unit\\
\hline
$C$&1&$\mu$F/cm$^2$&$E_4$&15&mV\\

$I_\text{ext}$&0.8&$\mu$A/cm$^2$&$E_\text{thresh}$&15&mV\\

$g_\text{L}$&0.005&$\mu$S/cm$^2$&$E_\text{slope}$&2&mV\\

$g_\text{Ca}$&0.015&$\mu$S/cm$^2$&$\phi_{N}$&0.0005&msec$^{-1}$\\

$g_\text{K}$&0.02&$\mu$S/cm$^2$&$L_0$&as in Figure&cm\\

$g_\text{syn}^\text{CPG}$&0.005&$\mu$S/cm$^2$&$L_\text{slope}$&as in Figure&cm\\

$g_\text{syn}^\text{FB}$&0.001&$\mu$S/cm$^2$&$\kappa$&2&none\\

$E_\text{L}$&-50&mV&$F_\ell$&2&N\\

$E_\text{Ca}$&100&mV&$F_0$&10&N\\

$E_\text{K}$&-80&mV&$b$&4$\times$10$^3$&N$\cdot$msec/cm\\

$E_\text{syn}^\text{CPG}$&-80&mV&$\tau$&2.45&none\\

$E_\text{syn}^\text{FB}$&$\pm80$&mV&$g$&2&none\\

$E_1$&0&mV&$a_0$&0.165&none\\

$E_2$&15&mV&$\beta$&0.703&none\\

$E_3$&0&mV\\
\hline
\end{tabular}
\end{table}

\section{Markin model details}
\label{app:Markin_model_details}

Based on \cite{markin2010afferent} and \cite{spardy2011dynamical}, the following provides modeling details not specified in Section \ref{sec:Markin model application}, especially on the biomechanics and feedback dynamics.
The limb motion is described by a
second-order differential equation:
\begin{equation*}
    I\ddot{q}=\frac{1}{2}mgl_s\cos{q}-b\dot{q}+M_F(q,\dot{q},V_\text{Mn-F},t)+M_E(\pi-q,-\dot{q},V_\text{Mn-E},t)+M_\text{GR}(q),
\end{equation*}
where $I=ml_s^2/3$ is the moment of inertia of the limb with respect to the suspension point and $b$ is the angular viscosity in the hinge joint. 
The first term accounts for the moment of the gravitational force; $M_F$ and $M_E$ are the moments of the muscle forces (see below); $M_\text{GR}$ represents the moment of the ground reaction force which is active only during the stance phase, given by 
\begin{align*}
M_\text{GR}(q)=\left\{\begin{aligned}
&-M_\text{GRmax}\cos{(q-\kappa)},\quad&\dot{q}\geq0\text{ (stance / power stroke)},\\
&0,\quad&\dot{q}<0\text{ (swing / recovery)}.
\end{aligned}\right.
\end{align*}
The parameter $\kappa$ describes the slope where the limb stands on, which is considered as the load parameter subjected to perturbations for this model. 
This parameter is not included in the original model \cite{markin2010afferent}; in effect in the original model $\kappa\equiv 0$.

In the muscle model, the muscle length is calculated as
\begin{equation*}
    L_F=\sqrt{a_1^2+a_2^2-2a_1a_2\cos{q}},\quad L_E=\sqrt{a_1^2+a_2^2-2a_1a_2\cos{(\pi-q)}},
\end{equation*}
and the moment arm is given by
\begin{equation*}
    h_F=(a_1a_2\sin{q})/L_F,\quad h_E=(a_1a_2\sin{(\pi-q)})/L_E,
\end{equation*}
for the flexor msucle and extensor muscle, respectively. 
Muscle velocities are defined as
\begin{equation*}
    v_F^m=vh_F,\quad v_E=-vh_E,
\end{equation*}
where $v=\dot{q}$ denotes the limb angular velocity.
The total force in each muscle follows the Hill-type model:
\begin{equation*}
    F=F_\text{max}(f(V)F_lF_v+F_p),
\end{equation*}
Constant $F_\text{max}$ is the maximal isometric force; $f(V)$ is the output of the corresponding motorneuron. 
The force–length dependence $F_l$ is given by\footnote{Careful investigation on the code provided by Markin revealed a typo in the function $F_l$ in the original papers. Here is the correct version.}
$$F_l=\exp{\left(-\left|\frac{l^\beta-1}{w}\right|\right)^\rho},$$
where $l$ is the normalized muscle length corresponding to $F_l = 1$, i.e. $l = L/L_\text{opt}$.
The force–velocity dependence $F_v$ is given by
\begin{align*}
F_v=\left\{\begin{aligned}
&\frac{b_1-c_1v^m}{v^m+b_1},\quad&v^m<0,\\
&\frac{b_2-c_2(l)v^m}{v^m+b_2},\quad&v^m\geq0,
\end{aligned}\right.
\end{align*}
where $c_2(l)=-5.34l^2+8.41l-4.7$.
The passive force $F_p$ is calculated as follows:
\begin{align*}
F_p=3.5\ln{\left(\exp{\left(\frac{l-1.4}{0.05}\right)}+1\right)}-0.02\left(\exp{\left(-18.7(l-0.79)\right)}-1\right).
\end{align*}
Given the above, the muscle moments are given by $M_F=F_Fh_F$ and $M_E=-F_Eh_E$.

To complete the model, the afferent feedback terms fed into the CPG neurons are in the form
\begin{align*}
    \text{Ia}&=k_\text{v\text{\uppercase\expandafter{\romannumeral1}a}}\text{sign}(v^m)|v_\text{norm}|^{\rho_v}+k_\text{dIa}d_\text{norm}+k_\text{nIa}f(V_\text{Mn})+C_\text{\text{Ia}},\\
    \text{II}&=k_\text{dII}d_\text{norm}+k_\text{nII}f(V_\text{Mn})+C_\text{II},\\
    \text{Ib}&=k_\text{FIb}F_\text{norm}.
\end{align*}
Here, $v_\text{norm}$ denotes the normalized muscle velocity ($v^m/L_\text{th}$); $d_\text{norm}$ denotes the normalized muscle length ($(L-L_\text{th})/L_\text{th}$ if $L\geq L_\text{th}$ and 0 otherwise); $F_\text{norm}$ denotes the normalized muscle force ($(F-F_\text{th})/F_\text{max}$ if $F\geq F_\text{th}$ and 0 otherwise).
The other functions in the CPG model are
\begin{align*}
m_\text{K}(V)&=1/\left(1+\exp{\left(-\frac{V+44.5}{5}\right)}\right),\\
m_\text{NaP}(V)&=1/\left(1+\exp{\left(-\frac{V+47.1}{3.1}\right)}\right),\\
h_\infty(V)&=1/\left(1+\exp{\left(\frac{V+51}{4}\right)}\right),\\
\tau_h(V)&=600/\cosh{\left(\frac{V+51}{8}\right)}.
\end{align*}
The values of the weight parameters for synaptic connections are provided in Table~\ref{table:Markin_weights}, and the other parameter values are listsed in Table~\ref{table:Markin_parameters}.
Simulation codes required to produce each figure are available at \texttt{https://github.com/zhuojunyu-appliedmath/Powerstroke-recovery}.

\begin{table}
\caption{Synaptic connection weights of the Markin model. The first row denotes the target neurons and the first column denotes the sources.}
\label{table:Markin_weights}
\centering
\begin{tabular}{lllllllllll}
\hline
&RG-F&RG-E&In-F&In-E&PF-F&PF-E&Mn-F&Mn-E&Int&Inab-E\\
\hline
\textbf{Drive connections,} $\mathbf{c_i}$&\\
Supra-spinal drive, $d$&0.08&0.08&&&0.40&0.40\\
External drive, $d_\text{Int}$&&&&&&&&&0.18\\
\hline
\textbf{Excitatory connections,} $\mathbf{a_{ji}}$\\
RG-F&&&0.41&&0.70\\
RG-E&&&&0.41&&0.70\\
PF-F&&&&&&&1.95\\
PF-E&&&&&&&&1.30&&0.35\\
Inab-E&&&&&&&&0.82\\
\hline
\textbf{Inhibitory connections,} $\mathbf{b_{ji}}$\\
In-F&&2.20&&&&6.60\\
In-E&2.20&&&&6.60&&&&2.80\\
Int&&&&&&&&&&0.55\\
\hline
\textbf{Feedback connections,} $\mathbf{w_{ki}}$\\
Ia-F&0.06&&0.27&&0.19\\
II-F&0.0348&&0.1566&&0.1102\\
Ia-E&&0.06&&0.44&&0.10&&&&0.16\\
Ib-E&&0.066&&0.484&&0.11&&&&0.176\\
\hline
\end{tabular}
\end{table}

\begin{table}[htb]
\caption{Parameter values of the Markin model}
\label{table:Markin_parameters}
\centering
\begin{tabular}{llllll}
\hline
Parameter&Value&Unit&Parameter&Value&Unit\\
\hline
$C$&20&pF&$M_\text{GRmax}$&585&N$\cdot$mm\\
$E_\text{Na}$&55&mV&$a_1$&60&mm\\
$E_\text{K}$&-80&mV&$a_2$&7&mm\\
$E_\text{L}$ for RG, PF, Mn&-64&mV&$F_\text{max}$ for flexor&72.5&N\\
$E_\text{L}$ for others&-60&mV&$F_\text{max}$ for extensor&37.7&N\\
$E_\text{synE}$&-10&mV&$\beta$&2.3&none\\
$E_\text{synI}$&-70&mV&$w$&1.6&none\\
$g_\text{NaP}$ for RG&3.5&nS&$\rho$&1.62&none\\
$g_\text{NaP}$ for PF&0.5&nS&$L_\text{opt}$&68&mm\\
$g_\text{NaP}$ for Mn&0.3&nS&$b_1$&-0.69&none\\
$g_\text{K}$&4.5&nS&$b_2$&0.18&none\\
$g_\text{L}$&1.6&nS&$c_1$&0.17&none\\
$g_\text{synE}$&10&nS&$k_\text{vIa}$&6.2&none\\
$g_\text{synI}$&10&nS&$k_\text{dIa}$&2&none\\
$V_{1/2}$&-30&mV&$k_\text{nIa}$&0.06&none\\
$V_\text{th}$&-50&mV&$C_\text{Ia}$&0.026&none\\
$k$ for Mn&3&mV&$k_\text{dII}$&1.5&none\\
$k$ for others&8&mV&$k_\text{nII}$&0.06&none\\
$d$&1.4&none&$C_\text{II}$&0&none\\
$s_k$&as in Figure&none&$k_\text{FIb}$&1&none\\
$m$&300&g&$\rho_v$&0.6&none\\
$g$&0.00981&\text{mm}/\text{ms}$^2$&$L_\text{th}$ for Ia&60.007&mm\\
$l_s$&300&mm&$L_\text{th}$ for II&58.457&mm\\
$b$&0.002&g$\cdot$mm$^2$/ms&$F_\text{th}$&3.393&N\\
\hline
\end{tabular}
\end{table}

\section{Variational analysis}
\label{app:variational_analysis}

The tools in the variational analysis were developed in \cite{wang2021shape} and generalized in \cite{yu2022homeostasis} and \cite{yu2023sensitivity}. 
Consider a parameterized continuous-time dynamical system defined on a domain $\Omega\subset\mathbb{R}^n$,
\begin{equation}
\label{eq:general_system_app}
\frac{d\mbz}{dt}=\mbF_\kappa(\mbz),
\end{equation}
where $\mbz\in\Omega$, $\kappa\in\mathcal{I}\subset\mathbb{R}$, and $\mbF_\kappa(\cdot)$ is either smooth or piecewise smooth in both $\mbz$ and $\kappa$.
Suppose $\mbF_\kappa(\cdot)$ admits a family of hyperbolic and asymptotically attracting limit cycles for $\kappa\in\mathcal{I}$ including $\kappa_0$, which we consider as the unperturbed limit cycle $\gamma_0(t)$.
We assume that the domain is partitioned into two subdomains $\Omega=R^\text{I}\cup R^\text{II}$ with the subdomain boundaries transverse to the flow of the unperturbed limit cycle, and that $\mbF_\kappa(\cdot)$ is smooth within each subdomain.

\subsection{Local timing response cuvre (lTRC)}
\label{subapp:lTRC}

For $\mbz\in R^i$, $i=\text{I}, \text{II}$, let $\Gamma^i(\mbz)$ be the time remaining until the unperturbed trajectory beginning at $\mbz$ exits the region.
By construction, along $\gamma_0$,
$$\frac{d\Gamma^i}{dt}(\mbz)=-1,\quad \mbz\in R^\text{i}.$$
Hence, by the chain rule
\begin{equation}
\label{eq:lTRC_chain_rule}
    \mbF_0(\mbz)\cdot\nabla\Gamma^i(\mbz)=-1,
\end{equation}
where we write $\mbF_0(\cdot)$ for $\mbF_{\kappa_0}(\cdot)$.
Define the \emph{local timing response curve} (lTRC) for subdomain $R^i$ to be $\eta^i(t):=\nabla\Gamma^i(\mbz(t))$.
Differentiating both sides of \eqref{eq:lTRC_chain_rule} with respect to $t$ yields 
\begin{equation}
\label{eq:app_lTRC_def}
\frac{d\eta^i}{dt}=-D\mbF_0(\gamma_0(t))^\intercal\eta^i,
\end{equation}
where $D\mbF_0$ denotes the Jacobian of $\mbF_0$.
Let $\mbz_0^{\text{out},i}$ denote the exit point of tracjectory $\gamma_0$ from region $R^i$ and $n^{\text{out},i}$ denote a normal vector of the exit boundary of $R^i$ at $\mbz_0^{\text{out},i}$.
Following \eqref{eq:lTRC_chain_rule},
$$\mbF_0(\mbz_0^{\text{out},i})\cdot\nabla\Gamma^i(\mbz_0^{\text{out},i})=-1,$$
which gives the boundary condition of $\eta^i$,
\begin{equation}
\label{eq:app_lTRC_boundary}
\eta^i(\mbz_0^{\text{out},i})=-\frac{n^{\text{out},i}}{(n^{\text{out},i})^\intercal\mbF_0(\mbz_0^{\text{out},i})}.
\end{equation}
The adjoint equation \eqref{eq:app_lTRC_def} together with the boundary condition  \eqref{eq:app_lTRC_boundary} defines the lTRC within region $R^i$.

One application of the lTRC is to calculate the duration the trajectory spent in each region (phase).
For a small positive $\epsilon$, consider $\kappa_\epsilon=\kappa_0+\epsilon\in\mathcal{I}$ and the corresponding perturbed trajectory $\gamma_\epsilon.$
We assume that 
$$T^i_\epsilon=T^i_0+\epsilon T_1^i+O(\epsilon^2),$$
where $T_0^i$ and $T_\epsilon^i$ represent the duration of the unperturbed trajectory and the perturbed trajectory spent in phase $i$, respectively, and $T_1^i$ is therefore the linear shift in the phase duration in response to the perturbation.
In \cite{yu2023sensitivity}, we provide a general expression for $T_1^i$:
\begin{equation}
\label{eq:T1_app}
T_1^i=\eta^i(\mbz_0^{\text{in},i})\cdot\frac{\partial\mbz_\kappa^{\text{in},i}}{\partial\kappa}\bigg|_{\epsilon=0}-\eta^i(\mbz_0^{\text{out},i})\cdot\frac{\partial\mbz_\kappa^{\text{out},i}}{\partial\kappa}\bigg|_{\epsilon=0}+\int_{t^{\text{in},i}}^{t^{\text{out},i}}\eta^i(\gamma_0(t))\cdot\frac{\partial\mbF_\kappa(\gamma_0(t))}{\partial\kappa}\bigg|_{\epsilon=0}\,dt.
\end{equation}
Here, at time $t^{\text{in},i}$ the unperturbed limit cycle $\gamma_0$ enters phase $i$ and at time $t^{\text{out},i}$ it exits the phase; $\mbz_\kappa^{\text{in},i}$ and $\mbz_\kappa^{\text{out},i}$ represent the entry point to and exit point from the phase $i$, respectively, for the limit cycle trajectory with $\kappa$.
The expression shows that the first-order shift in the phase duration consists of three terms: the first term accounts for the impact of the perturbation on the entry point to the region; the second term arises from the impact on the exit point; the integral term shows the impact on the vector field during the transit from ingress to egress.

\subsection{Infinitesimal shape response curve  (iSRC)}
\label{subapp:iSRC}

Consider the perturbed trajectory $\gamma_\epsilon$ and the unperturbed trajectory $\gamma_0$.
Within each subdomain $R^i$, we expand $\gamma_\epsilon$:
$$\gamma_\epsilon(\tau^i_\epsilon(t))=\gamma_0(t)+\epsilon\gamma_1(t)+O(\epsilon^2),$$
where $\tau^i_\epsilon(t)$ is introduced as a rescaled time coordinate so that the expansion is uniform with respect to $t$. 
It satisfies
$$\tau^i_\epsilon(0)=0,\quad \tau^i_\epsilon(t+T_0^i)=\tau^i_\epsilon(t)+T^i_\epsilon$$
The vector function $\gamma_1(t)$ is defined as the \emph{infinitesimal shape response curve} (iSRC), which is piecewise-specified with period $T_0$.
In region $R^i$, the iSRC satisfies a nonhomogenous variational equation
\begin{equation}
\label{eq:iSRC_app}
    \frac{d\gamma_1^i(t)}{dt}=D\mbF_0(\gamma_0(t))\gamma_1^i(t)+\nu_1^i(t)\mbF_0(\gamma_0(t))+\frac{\partial\mbF_\kappa(\gamma_0(t))}{\partial\kappa}\bigg|_{\epsilon=0},
\end{equation}
where $\nu_1^i=\frac{\partial^2\tau^i_\epsilon(t)}{\partial\epsilon\partial t}\big|_{\epsilon=0}$ measures the (local) timing sensitivity to the perturbation in region $R^i$.
In the special case of $\tau^i_\epsilon$ with a linear scaling, i.e., $\tau_\epsilon^i(t)=(T_\epsilon^i/T_0^i)t$, we have $\nu_1^i=T_1^i/T_0^i$ independent of $t$, where $T_1^i$ is given by \eqref{eq:T1_app}.

The iSRC equation \eqref{eq:iSRC_app} requires an initial condition, which is
$$\gamma_1^i(t^{\text{in},i})=\lim_{\epsilon\rightarrow0}\frac{\mbp_\epsilon^i-\mbp_0^i}{\epsilon},$$
where $\mbp_\epsilon^i$ and $\mbp_0^i$ represent the intersection points of the limit cycle trajectories with the boundary surfaces between regions.
Generally, changing the surfaces (initial conditions) results in the consequent iSRC functions related by a simple phase shift, which has no effect on the sensitivity given by \eqref{eq:sensitivity_formula1}.
See Lemma 1 in \cite{yu2022homeostasis} or Lemma 2.3 in \cite{wang2021shape}) for proof.

\subsection{Derivation of sensitivity formula \eqref{eq:sensitivity_formula1}}
\label{subapp:sensitivity_formula1}

This section presents the derivation of equation~\eqref{eq:sensitivity_formula1}.
It is a general formula to calculate the sensitivity of an averaged quantity with respect to the change in any parameter, applicable to the powerstroke-recovery systems.
The derivation follows the same exposition given in \cite{yu2022homeostasis} but with some modifications.

\begin{figure}[htb]
\centering
\includegraphics[width=8cm]{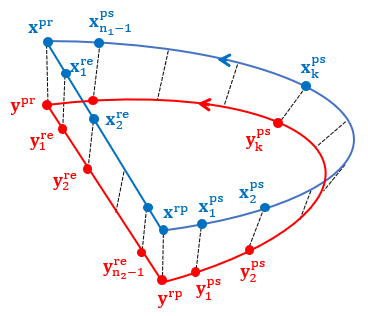}%
\caption{\label{fig:points_discretization} Discretization of limit cycles by points comprising the limit cycles' orbits satisfying \eqref{eq:general_system_app}, illustrating the construction for comparing the average of a quantity around the two orbits. 
Figure modified from \cite{yu2022homeostasis}.
The loop $\mbx$ (blue trace) represents the unperturbed limit cycle with the default parameter value $\kappa_0$; the loop $\mby$ (red trace) represents the perturbed limit cycle with $\kappa_\epsilon=\kappa_0+\epsilon$.
Each orbit consists of two phases --- powerstroke (ps) and recovery (re).
During the powerstroke, each orbit is divided into $n_1$ steps of equal time $T^\text{ps}_x/n_1$ or $T^\text{ps}_y/n_1$, respectively, giving points ($\mbx^\text{ps}_1, \mbx^\text{ps}_2,\cdots,\mbx^\text{ps}_{n_1-1}$) and ($\mby^\text{ps}_1, \mby^\text{ps}_2,\cdots,\mby^\text{ps}_{n_1-1}$).
During the recovery, each orbit is divided into $n_2$ steps of equal time $T_x^\text{re} /n_2$ or $T_y^\text{re} /n_2$, respectively, giving points ($\mbx^\text{re}_1, \mbx^\text{re}_2,\cdots,\mbx^\text{re}_{n_2-1}$) and ($\mby^\text{re}_1, \mby^\text{re}_2,\cdots,\mby^\text{re}_{n_2-1}$).
The phase transition points are marked as $\mbx^{\text{rp}},\mbx^{\text{pr}},\mby^{\text{rp}},\mby^{\text{pr}}$ correspondingly.}
\end{figure}

Denote the unperturbed limit cycle (with $\kappa_0$) by $\mbx$ and the perturbed limit cycle (with $\kappa_\epsilon=\kappa_0+\epsilon$) by $\mby$, respectively.
Denote the periods by
$$T_x=T_x^\text{ps}+T_x^\text{re},\qquad T_y=T_y^\text{ps}+T_y^\text{re}.$$
During the powerstroke, each limit cycle orbit is divided into $n_1$ steps of equal time $T_x^\text{ps}/n_1$ or $T_y^\text{ps}/n_1$; during the recovery, each orbit is divided into $n_2$ steps of equal time $T_x^\text{re}/n_2$ or $T_y^\text{re}/n_2$. 
Let $n=n_1+n_2.$
Mark off the points as in Figure~\ref{fig:points_discretization}.
Then,
\begin{align*}
    &\mbx_k^\text{ps}=\mbx\left(\frac{k}{n_1}T^\text{ps}_x\right),\quad\mby_k^\text{ps}=\mby\left(\frac{k}{n_1}T^\text{ps}_y\right),\quad k=0,1,\cdots,n_1,\\
    &\mbx_k^\text{re}=\mbx\left(T^\text{ps}_x+\frac{k}{n_2}T^\text{re}_x\right),\quad\mby_k^\text{re}=\mby\left(T^\text{ps}_y+\frac{k}{n_2}T^\text{re}_y\right),\quad k=0,1,\cdots,n_2.
\end{align*}
Note that at the powerstroke-to-recovery (pr) transition state,
\begin{align*}
    \mbx^{\text{pr}}=\mbx^\text{ps}_{n_1}=\mbx^\text{re}_{0},\qquad\mby^{\text{pr}}=\mby^\text{ps}_{n_1}=\mby^\text{re}_{0},
\end{align*}
and at the recovery-to-powerstroke (rp) transition state,
\begin{align*}
    \mbx^{\text{rp}}=\mbx^\text{ps}_0=\mbx^\text{re}_{n_2},\qquad\mby^{\text{rp}}=\mby^\text{ps}_0=\mby^\text{re}_{n_2}.
\end{align*}
Let $Q_x$ be the average of the quantity $q(\mbx)$ around the trajectory $\mbx$, i.e.,
\begin{equation}
\label{eq:average1_for_x}
    Q_x=\frac{1}{T_x}\int_0^{T_x}q(\mbx(t))\,dt=\frac{1}{T_x}\left[\int_0^{T_x^\text{ps}}q(\mbx(t))\,dt+\int_{T_x^\text{ps}}^{T_x}q(\mbx(t))\,dt\right],
\end{equation}
and similarly $Q_y$. 
Suppose $q$ and the limit cycles are piecewise smooth, where the discontinuity occurs at the transitions between the two phases.
As $n_1,n_2\rightarrow\infty$, following \eqref{eq:average1_for_x}, since the set of discontinuous points has measure zero, we have
\begin{align*}
    Q_x&=\frac{1}{T_x}\left[\frac{T_x^\text{ps}}{n_1}\sum_{k=0}^{n_1-1}q(\mbx_k^\text{ps})+\frac{T_x^\text{re}}{n_2}\sum_{k=0}^{n_2-1}q(\mbx_k^\text{re})\right]+O\left(\frac{1}{n^2}\right)\\
    &=\frac{1}{n_1}\sum_{k=0}^{n_1-1}\frac{T_x^\text{ps}}{T_x}q(\mbx_k^\text{ps})+\frac{1}{n_2}\sum_{k=0}^{n_2-1}\frac{T_x^\text{re}}{T_x}q(\mbx_k^\text{re})+O\left(\frac{1}{n^2}\right),
\end{align*}
with a similar expression for $Q_y$.
Then,
\begin{align*}
    Q_y-Q_x=\frac{1}{n_1}\sum_{k=0}^{n_1-1}\left(\frac{T_y^\text{ps}}{T_y}q(\mby_k^\text{ps})-\frac{T_x^\text{ps}}{T_x}q(\mbx_k^\text{ps})\right)+\frac{1}{n_2}\sum_{k=0}^{n_2-1}\left(\frac{T_y^\text{re}}{T_y}q(\mby_k^\text{re})-\frac{T_x^\text{re}}{T_x}q(\mbx_k^\text{re})\right)+O\left(\frac{1}{n^2}\right).
\end{align*}
Let the fraction of the powerstroke duration be $\beta$, i.e.,
$$\beta=\frac{T^\text{ps}}{T},\qquad1-\beta=\frac{T^\text{re}}{T}.$$
Expand $\beta_y$ and $q(\mby_k)$ around $\epsilon=0$:
\begin{align*}
    \beta_y&=\beta_x+\epsilon\beta_1+O(\mu^2),\\
    q(\mby^\text{ps}_k)&=q(\mbx^\text{ps}_k)+\nabla q(\mbx^\text{ps}_k)\cdot(\mby^\text{ps}_k-\mbx^\text{ps}_k)+\epsilon\frac{\partial q(\mbx_k^\text{ps})}{\partial\kappa}+O(\epsilon^2),\quad k=1,2,\cdots n_{1}-1,\\
    q(\mby^\text{re}_k)&=q(\mbx^\text{re}_k)+\nabla q(\mbx^\text{re}_k)\cdot(\mby^\text{re}_k-\mbx^\text{re}_k)+O(\epsilon^2),\quad k=1,2,\cdots n_{2}-1,
\end{align*}
where $\beta_1$ represents the linear shift in $\beta^x$ under the perturbation. 
Since the quantity $q$ may be explicitly affected by the perturbation during the powerstroke, then there is an additional term $\frac{\partial q}{\partial\kappa}$ in the expansion of $q(\mby^\text{ps}_k)$.
Note that the gradient $\nabla q$ is not well-defined at the discontinuous points.
Then,
\begin{align*}
    Q_y-Q_x=&\frac{1}{n_1}\sum_{k=1}^{n_1-1}\left[\beta^yq(\mby_k^\text{ps})-\beta^xq(\mbx_k^\text{ps})\right]+\frac{1}{n_2}\sum_{k=1}^{n_2-1}\left[(1-\beta^y)q(\mby_k^\text{re})-(1-\beta^x)q(\mbx_k^\text{re})\right]\\
    &\qquad+\frac{1}{n_1}\left[\beta^yq(\mby^\text{rp})-\beta^xq(\mbx^\text{rp})\right]+\frac{1}{n_2}\left[(1-\beta^y)q(\mby^\text{pr})-(1-\beta^x)q(\mbx^\text{pr})\right]+O\left(\frac{1}{n^2}\right)\\
    =&\frac{1}{n_1}\sum_{k=1}^{n_1-1}\left[\beta^x\nabla q(\mbx_k^\text{ps})\cdot(\mby^\text{ps}_k-\mbx^\text{ps}_k)+\epsilon\beta^x\frac{\partial q(\mbx_k^\text{ps})}{\partial\kappa}+\epsilon\beta_1q(\mbx^\text{ps}_k)\right]\\
    &+\frac{1}{n_2}\sum_{k=1}^{n_2-1}\left[(1-\beta^x)\nabla q(\mbx_k^\text{re})\cdot(\mby^\text{re}_k-\mbx^\text{re}_k)-\epsilon\beta_1q(\mbx^\text{re}_k)\right]\\
    &+\frac{1}{n_1}\left[\beta^x(q(\mby^\text{rp})-q(\mbx^\text{rp}))+\epsilon\beta_1q(\mby^\text{rp})\right]\\
    &+\frac{1}{n_2}\left[(1-\beta^x)(q(\mby^\text{pr})-q(\mbx^\text{pr}))-\epsilon\beta_1q(\mby^\text{pr})\right]+O(\epsilon^2)+O\left(\frac{1}{n^2}\right).
\end{align*}
Using the iSRC $\gamma_1$ of $\mbx$, we obtain
\begin{align*}
    Q_y-Q_x=&\frac{\epsilon}{n_1}\sum_{k=1}^{n_1-1}\left[\beta^x\nabla q(\mbx_k^\text{ps})\cdot\gamma_1(\mbx^\text{ps}_k)+\beta^x\frac{\partial q(\mbx_k^\text{ps})}{\partial\kappa}+\beta_1q(\mbx^\text{ps}_k)\right]\\
    &+\frac{\epsilon}{n_2}\sum_{k=1}^{n_2-1}\left[(1-\beta^x)\nabla q(\mbx_k^\text{re})\cdot\gamma_1(\mbx^\text{re}_k)-\beta_1q(\mbx^\text{re}_k)\right]\\
    &+\frac{1}{n_1}\left[\beta^x(q(\mby^\text{rp})-q(\mbx^\text{rp}))+\epsilon\beta_1q(\mby^\text{rp})\right]\\
    &+\frac{1}{n_2}\left[(1-\beta^x)(q(\mby^\text{pr})-q(\mbx^\text{pr}))-\epsilon\beta_1q(\mby^\text{pr})\right]+O(\epsilon^2)+O\left(\frac{1}{n^2}\right).
\end{align*}
Dividing both sides by $\epsilon$ yields
\begin{align*}
    \frac{Q_y-Q_x}{\epsilon}=&\frac{1}{n_1}\sum_{k=1}^{n_1-1}\left[\beta^x\nabla q(\mbx_k^\text{ps})\cdot\gamma_1(\mbx^\text{ps}_k)+\beta^x\frac{\partial q(\mbx_k^\text{ps})}{\partial\kappa}+\beta_1q(\mbx^\text{ps}_k)\right]\\
    &+\frac{1}{n_2}\sum_{k=1}^{n_2-1}\left[(1-\beta^x)\nabla q(\mbx_k^\text{re})\cdot\gamma_1(\mbx^\text{re}_k)-\beta_1q(\mbx^\text{re}_k)\right]\\
    &+\frac{\beta^x}{\epsilon n_1}\left(q(\mby^\text{rp})-q(\mbx^\text{rp})\right)+\frac{\beta_1}{n_1}q(\mby^\text{rp})\\
    &+\frac{1-\beta^x}{\epsilon n_2}\left(q(\mby^\text{pr})-q(\mbx^\text{pr})\right)-\frac{\beta_1}{n_2}q(\mby^\text{pr})+O(\epsilon)+O\left(\frac{1}{n^2}\right).
\end{align*}
Taking the limits $\epsilon\rightarrow0$ and $n_1,n_2\rightarrow\infty$, we obtain 
\begin{align*}
    \frac{\partial Q}{\partial\kappa}(\kappa_0)=&\frac{1}{T_x^\text{ps}}\int_0^{T^\text{ps}_x}\left[\beta^x\nabla q(\mbx(t))\cdot\gamma_1(t)+\beta^x\frac{\partial q(\mbx_k^\text{ps})}{\partial\kappa}+\beta_1q(\mbx(t))\right]\,dt\\
    &+\frac{1}{T^\text{re}_x}\int_{T_x^\text{ps}}^{T_x}\left[(1-\beta^x)\nabla q(\mbx(t))\cdot\gamma_1(t)-\beta_1q(\mbx(t))\right]\,dt\\
    &+\lim_{\substack{\epsilon\rightarrow0\\n_1\rightarrow\infty}}\frac{\beta^x}{\epsilon n_1}\left(q(\mby^\text{rp})-q(\mbx^\text{rp})\right)+\lim_{\substack{\epsilon\rightarrow0\\n_2\rightarrow\infty}}\frac{1-\beta^x}{\epsilon n_2}\left(q(\mby^\text{pr})-q(\mbx^\text{pr})\right).
\end{align*}
For the last two terms, use the directional derivative in the direction (say $\mbu$) tangent to the recovery-powerstroke boundary, and expand $q(\mby^\text{rp})$ and $q(\mby^\text{pr})$ to be
\begin{align*}
    q(\mby^\text{rp})&=q(\mbx^\text{rp})+\epsilon\nabla_\mbu q(\mbx^\text{rp})+O(\epsilon^2)\\
    q(\mby^\text{pr})&=q(\mbx^\text{pr})+\epsilon\nabla_\mbu q(\mbx^\text{pr})+O(\epsilon^2)
\end{align*}
Then,
\begin{align*}
    \lim_{\substack{\epsilon\rightarrow0\\n_1\rightarrow\infty}}\frac{1}{\epsilon n_1}\left(q(\mby^\text{rp})-q(\mbx^\text{rp})\right)&=\lim_{\substack{\epsilon\rightarrow0\\n_1\rightarrow\infty}}\frac{1}{\epsilon n_1}\left(\epsilon\nabla_\mbu q(\mbx^\text{rp})+O(\epsilon^2)\right)\\
    &=\lim_{n_1\rightarrow\infty}\frac{\nabla_\mbu q(\mbx^\text{rp})}{n_1}=0,\\
    \lim_{\substack{\epsilon\rightarrow0\\n_2\rightarrow\infty}}\frac{1}{\epsilon n_2}\left(q(\mby^\text{pr})-q(\mbx^\text{pr})\right)&=\lim_{n_2\rightarrow\infty}\frac{\nabla_\mbu q(\mbx^\text{pr})}{n_2}=0.
\end{align*}
Therefore, we obtain the sensitivity of the average $Q$ for the unperturbed system:
\begin{equation}
\label{eq:sensitivity_formula1_app}
\begin{split}
    \frac{\partial Q}{\partial\kappa}(\kappa_0)&=\frac{1}{T_0^\text{ps}}\int_0^{T_0^\text{ps}}\left[\beta_0\left(\nabla q_{0}(\gamma_0(t))\cdot\gamma_1(t)+\frac{\partial q_\kappa(\gamma_0(t))}{\partial\kappa}\bigg|_{\epsilon=0}\right)+\beta_1q_0(\gamma_0(t))\right]\,dt\\
    &\qquad\qquad+\frac{1}{T^\text{re}_0}\int_{T_0^\text{ps}}^{T_0}\left[(1-\beta_0)\nabla q_0(\gamma_0(t))\cdot\gamma_1(t)-\beta_1q_0(\gamma_0(t))\right]\,dt.
\end{split}
\end{equation}
Note that when $q$ represents the change rate of progress, then for the models considered here (and for Shaw et al.'s \textit{Aplysia} feeding model \citep{shaw2015significance,lyttle2017robustness,wang2022variational}) $q$ vanishes during the recovery, indicating that $q_0(\gamma_0(t))\equiv0$ for $t\in(T_0^\text{ps},T_0]$ and that the second integral in \eqref{eq:sensitivity_formula1_app} reduces to zero. 
Formula \eqref{eq:sensitivity_formula1} is derived.

\bibliography{refs}
\bibliographystyle{apalike}

\end{document}